\documentclass[aps,prb,superscriptaddress,floatfix,amsmath,amssymb,reprint]{revtex4-2}

\pdfoutput=1

\usepackage{color}
\usepackage{frcursive}
\usepackage{bbm}
\usepackage{mathtools}
\usepackage{graphicx}
\usepackage{enumitem}
\usepackage{bbm}
\usepackage{relsize}
\usepackage{xcolor}
\usepackage[caption=false]{subfig}
\usepackage{amsmath}	%American Mathematical Society math package
\usepackage{amssymb}    %American Mathematical Society symbol package
\usepackage[pdftex]{hyperref}

\begin{document}
		
		\title{Two-dimensional Dirac matter in the semiclassical regime}
		
		\author{Fran\c{c}ois Fillion-Gourdeau}
		\email{francois.fillion@inrs.ca}
		\affiliation{Institute for Quantum Computing, University of Waterloo, Waterloo, Ontario, Canada, N2L 3G1}
		\affiliation{Infinite Potential Laboratories, Waterloo, Ontario, Canada, N2L 0A9}
		
		\author{Emmanuel Lorin}
		\email{elorin@math.carleton.ca}
		\affiliation{School of Mathematics and Statistics, Carleton University, Ottawa, Canada}
		\affiliation{Centre de Recherches Math\'ematiques, Universit\'e de Montr\'eal, Montr\'eal, Canada.}
		
		\author{Steve MacLean}
		\email{steve.maclean@inrs.ca}
		
		\affiliation{Institute for Quantum Computing, University of Waterloo, Waterloo, Ontario, Canada, N2L 3G1}
		\affiliation{Infinite Potential Laboratories, Waterloo, Ontario, Canada, N2L 0A9}
		\affiliation{Universit\'{e} du Qu\'{e}bec, INRS-\'{E}nergie, Mat\'{e}riaux et T\'{e}l\'{e}communications, Varennes, Qu\'{e}bec, Canada J3X 1S2}
		
		\date{\today}

\begin{abstract}
The semiclassical regime of 2D static Dirac matter is obtained from the Dirac equation in curved space-time. To simplify the formulation, the Cartesian space-time geometry parametrization is transformed to isothermal coordinates using quasi-conformal transformations. Using this framework, it is demonstrated that to first order, the semiclassical approximation yields the relativistic Lorentz force equation with additional fictitious forces related to the space curvature and to the mass gradient. An effective graded index of refraction is defined and is directly linked to the metric component in isothermal coordinates.  Semiclassical trajectories are evaluated in a simple example by solving the equation of motion and the Beltrami equation numerically.
\end{abstract}

\maketitle

\section{Introduction}
The Dirac equation is at the core of our understanding of matter, being the generalization of the Schr\"{o}dinger equation for relativistic spin-1/2 particles. Originally, this equation was applied to the theoretical description of elementary particles like the electron or quarks, for high energy physics applications where relativistic effects are important. However, with the advent of some new materials and their effective low-energy description, it has been realized that the Dirac equation has an even more widespread field of application that also includes condensed matter physics. Dirac matter \cite{doi:10.1080/00018732.2014.927109,CAYSSOL2013760} encompasses all quantum systems theoretically described by the Dirac equation such as relativistic fermions, but now includes many quantum materials such as graphene \cite{PhysRev.71.622,doi:10.1142/9789814287005_0002,RevModPhys.81.109}, topological insulators \cite{RevModPhys.82.3045,RevModPhys.88.021004,RevModPhys.83.1057}, Dirac semimetals \cite{RevModPhys.90.015001}, some high-temperature supraconductors \cite{RevModPhys.78.373} and liquid Helium-3 \cite{volovik1992exotic}. It also includes exotic structures such as artificial graphene \cite{PhysRevLett.111.185307}, photonic graphene analogs \cite{PhysRevA.78.033834,PhysRevA.75.063813,PhysRevLett.100.113903,PhysRevLett.111.103901}, photonic topological insulators \cite{khanikaev2013photonic} and phononic metamaterials \cite{mousavi2015topologically}. Many of these materials are in the class of two-dimensional materials, where the underlying symmetry of the cristal lattice makes for a relativistic-like description of quasi-particles in the low-energy limit. 

Understanding the properties of such systems is a theoretical problem requiring explicit solutions to the Dirac equation. Numerical and analytical methods have been developed to achieve such a feat \cite{PhysRevA.59.604,SUCCI1993327,FILLIONGOURDEAU20121403,BAO2004663,HAMMER2014728,HAMMER201440,PhysRevE.96.053312,ANTOINE2020109412,PhysRevE.103.013312,PhysRevB.98.155419} but it remains a challenging task. Assuming the external potentials are smooth enough, an interesting alternative is to use the semiclassical approximation, in which the problem reduces to the solution of a classical-like system of equations \cite{maslov2001semi}. This approach has been implemented on the Dirac equation to understand and describe the relativistic quantum dynamics of fermions under various external fields \cite{maslov2001semi,BOLTE1999125,SPOHN2000420,doi:10.1063/1.1604455,REIJNDERS2013155,PhysRevLett.81.1987}. Motivated by applications such as the Dirac fermion microscope \cite{boggild2017two}, the semiclassical technique has also been applied to charge carriers in graphene \cite{PhysRevB.77.245413,REIJNDERS201865,PhysRevB.103.045404} for analyzing Veselago lenses \cite{PhysRevB.96.045305}. This gave rise to the field of ``electron optics'' in graphene, where ray-optics can be used to understand the behavior of electrons in Dirac materials.  

In comparison, less work has been performed on the semiclassical limit of the curved-space Dirac equation \cite{PhysRevD.71.064016,GOSSELIN2007356,doi:10.1142/S0217751X08040214,Montani2008}, possibly because the coupling to a gravitational field increases technical complications. Nevertheless, the curved-space Dirac equation is also relevant for the description of systems in condensed matter, such as strained graphene \cite{PhysRevB.87.165131,VOZMEDIANO2010109,PhysRevLett.108.227205,VOLOVIK2014352,AMORIM20161,OLIVALEYVA20152645,Naumis_2017} or more generally, for straintonics in Dirac materials. In this approach, electron control can be achieved by mechanical deformations, permitting the focusing \cite{Stegmann_2016,PhysRevE.103.013312} or confinement \cite{PhysRevB.98.155419} of charge carriers.    

In this article, the semiclassical limit of static 2D Dirac matter is investigated in the general case where the quantum system is described by the curved-space Dirac equation with a space-dependent mass gap. Thus, it provides a general framework to investigate the behavior of 2D fermions in numerous physical systems. To perform this task, isothermal coordinates are introduced to simplify the description significantly while making the physical interpretation more transparent. The physical significance of these particular coordinates will be emphasized by connecting them to a graded refractive index and electron optics.

This article is separated as follows. In Sec. \ref{sec:dirac_mat}, the Dirac equation in curved-space that models Dirac matter is written in isothermal coordinates. The semiclassical limit this equation is obtained in Sec. \ref{sec:semiclass}. In Sec. \ref{sec:ex}, we consider an explicit example for a Gaussian wave packet scattering on an out-of-plane deformation. Finally, we conclude in Sec. \ref{sec:conclu}.  

\section{Theoretical model of Dirac matter \label{sec:dirac_mat}}

Dirac matter shall be defined here as any 2D quantum systems whose fermionic degrees of freedom are characterized by the Dirac equation in curved-space. Therefore, this equation is reviewed in this section. In particular, we demonstrate that a simple expression for the curved-space Dirac equation can be obtained by using isothermal coordinates. 

\subsection{Curved-space Dirac equation}
In covariant form, the curved-space Dirac equation is given by \cite{PhysRevLett.108.227205,pollock2010dirac,PhysRevB.92.245110}
\begin{align}
\label{eq:dirac_cov}
\left[ \texttt{i}v_{D} \hbar \bar{\gamma}^{\mu}(q)D_{\mu} - m_{D}(q)v_{D}^{2} \right]\psi(q) = 0,
\end{align} 
where $\psi(q)$ is the two-component spinor wave function, $m_{D}$ is the mass of the fermion state, $v_{D}$ is the Dirac velocity, $q=(t,\boldsymbol{q})$ is a set of curvilinear coordinates (bold symbols are 2D vectors), $D_{\mu}=\partial_{\mu} + \Omega_{\mu} - {\tt i} A_{\mu}$ is the curved-space covariant derivative (that includes the spin connection $\Omega_{\mu}$ and the coupling to a vector field $A_{\mu}$) and $\bar{\gamma}(q) = ( \bar{\gamma}^{0}(q),\bar{\gamma}^{i}(q) )$ are the generalized gamma matrices. A curved-space manifold is associated to such systems via the generalized gamma matrices isomorphic to the local Clifford algebra
\begin{align}
\{\bar{\gamma}^{\mu}(q), \bar{\gamma}^{\nu}(q)\} = 2g^{\mu \nu} (q), 
\end{align}
and thus, can be written in terms of the vielbein as  $\bar{\gamma}^{\mu} = e_{a}^{\mu}(q) \gamma^{a}$, where $\gamma^{a}$ are the standard flat-space Dirac matrices. As usual, $g^{\mu \nu}(q)$ is the metric of the space-time manifold.

Specializing this formulation to static manifolds in Cartesian spatial coordinates ($q = (t,\boldsymbol{x})$), the most general metric yields a line element of the form
\begin{align}
\label{eq:metric_cart}
ds^{2} = v_{D}^{2}dt^{2} - E(\boldsymbol{x})dx^{2} - G(\boldsymbol{x}) dy^{2} -2F(\boldsymbol{x})dxdy,
\end{align} 
where $E,F,G$ are the components of the surface first fundamental form. This corresponds to the metric of a 2D surface embedded in a 3D Euclidean space written parametrically as $ \vec{R}(\boldsymbol{x}) = (X(\boldsymbol{x}),Y(\boldsymbol{x}), Z(\boldsymbol{x}))^{T}$. In terms of the surface parametrization, the metric components are given by
\begin{align}
E(\boldsymbol{x}) &= (\partial_{x}X)^{2} + (\partial_{x}Y)^{2} + (\partial_{x}Z)^{2}, \\
G(\boldsymbol{x}) &= (\partial_{y}X)^{2} + (\partial_{y}Y)^{2} + (\partial_{y}Z)^{2}, \\
F(\boldsymbol{x}) &= (\partial_{x}X)(\partial_{y}X) + (\partial_{x}Y)(\partial_{y}Y) + (\partial_{x}Z)(\partial_{y}Z).
\end{align}
It is possible to write the Dirac equation \eqref{eq:dirac_cov} explicitly in Cartesian coordinates for a general surface characterized by the metric \eqref{eq:metric_cart}. However, the resulting expression is intricate because both the metric and the vielbein are not diagonal \cite{PhysRevE.103.013312}. To simplify the analysis, we introduce a change of variables to isothermal coordinates.

\subsection{Isothermal coordinates}
It is convenient to introduce isothermal coordinates $\boldsymbol{r} = (r_{1},r_{2})$ to obtain an explicit expression for the Dirac equation. In these coordinates, the metric is given by \cite{PhysRevD.92.125005}:
\begin{align}
\label{eq:metric_iso}
ds^{2} = v_{F}^{2} dt^{2} - \rho(\boldsymbol{r}) d\boldsymbol{r}^{2}.
\end{align} 
In 2D, there always exists a change of coordinates $\boldsymbol{x} \rightarrow \boldsymbol{r}$ that diagonalizes the metric in Eq. \eqref{eq:metric_cart} and transforms it to Eq. \eqref{eq:metric_iso} \cite{ahlfors2006lectures}. %As seen below, this transformation is important for two main reasons: 1) the function $\rho$ can actually be interpreted as an effective diffractive index in the semiclassical limit and 2) the Dirac equation in curved-space time has a simpler form, allowing for the development of accurate numerical and approximation schemes \cite{PhysRevE.103.013312}. 
This mapping is given by a quasi-conformal transformation obtained by solving the Beltrami equation \cite{origami}:
\begin{align}
\label{eq:bel_eq}
P(\boldsymbol{x}) \nabla_{\boldsymbol{x}} r_{1}(\boldsymbol{x}) &= JP(\boldsymbol{x}) \nabla_{\boldsymbol{x}} r_{2}(\boldsymbol{x}),
\end{align}
where $J = -{\tt i} \sigma_{y}$ and $P$ is defined as
\begin{align}
P(\boldsymbol{x}) &= 
\frac{1}{\sqrt{1-|\mu(\boldsymbol{x})|^2}}
\left[ \begin{array}{cc}
1-\mu_{\mathrm{R}}(\boldsymbol{x}) & -\mu_{\mathrm{I}}(\boldsymbol{x}) \\
-\mu_{\mathrm{I}}(\boldsymbol{x})& 1+\mu_{\mathrm{R}}(\boldsymbol{x})
\end{array} \right] .
\end{align}
The function $P$ depends on the real ($\mu_{\mathrm{R}}$) and imaginary ($\mu_{\mathrm{I}}$) parts of the Beltrami coefficient:
\begin{align}
\label{eq:beltrami_coeff}
\mu(\boldsymbol{x}) = \frac{E(\boldsymbol{x}) - G(\boldsymbol{x}) + 2{\tt i}F(\boldsymbol{x})}{E(\boldsymbol{x}) + G(\boldsymbol{x}) + 2 \sqrt{E(\boldsymbol{x})G(\boldsymbol{x}) - F^{2}(\boldsymbol{x})}} < 1.
\end{align}
Isothermal coordinates can then be obtained by solving Eq. \eqref{eq:bel_eq} analytically or numerically while the metric diagonal component is given by
\begin{align}
\label{eq:rho}
\rho(\boldsymbol{x}) = \frac{E(\boldsymbol{x}) + G(\boldsymbol{x}) + 2 \sqrt{E(\boldsymbol{x})G(\boldsymbol{x}) - F^{2}(\boldsymbol{x})}}{\left[\partial_{x}r_{1}(\boldsymbol{x}) + \partial_{y}r_{2}(\boldsymbol{x})\right]^{2} + \left[\partial_{x}r_{2}(\boldsymbol{x}) - \partial_{y}r_{1}(\boldsymbol{x})\right]^{2}}.
\end{align}
Equation \eqref{eq:rho} actually connects the diagonal components of the metric in isothermal coordinates to the parametric surface via a solution of Eq. \eqref{eq:bel_eq}.

Armed with these coordinates, we now write Eq. \eqref{eq:dirac_cov} in a Schr\"{o}dinger-like form. Isothermal coordinates are critical to realize this step because they also make the vielbein diagonal.
For the metric in isothermal coordinates given in Eq. \eqref{eq:metric_iso}, the Dirac equation reads (see Appendix \ref{app:sch} and \cite{PhysRevE.103.013312} for details of the calculation):
\begin{align}
\label{eq:dirac_iso}
{\tt i}\hbar \partial_t \psi(t,{\boldsymbol r}) = \hat{H}
 \psi(t,{\boldsymbol r})  , \\
\hat{H} = 
-{\tt i}\frac{\hbar v_{D}}{\sqrt{\rho(\boldsymbol{r})}}\alpha^{i}   \tilde{D}_{i} + m_{D}(\boldsymbol{r})v_{D}^{2}\beta + V(\boldsymbol{r}) ,
\end{align}
where $V = A_{0}$ is the scalar potential and where a standard representation of Dirac matrices is chosen: $\alpha^{i} = \sigma^{i}$ (for $i=1,2$) and $\beta = \sigma^{3}$ ($\sigma^{i}$ are Pauli matrices). Furthermore, we have that $\tilde{D}_{i}=\partial_{i} + \tilde{\Omega}_{i} - {\tt i} A_{i}$ where the affine spin connection is given by
\begin{align}
\label{eq:spin_aff}
\tilde{\Omega}_{i}(\boldsymbol{r})
= -\frac{1}{4} \partial_{i} \ln \big(\rho(\boldsymbol{r}) \bigr),
\end{align}
which makes the Hamiltonian operator self-adjoint (see Appendix \ref{app:sch})), 
while the vector potential is split in two contributions $A_{i} = \frac{1}{\hbar}A^{(0)}_{i} + A^{(1)}_{i}$ with a different origin. $A^{(0)}_{i}$ is a usual vector potential that is provided by an external electromagnetic field, for example, and thus contributes to zeroth order in $\hbar$. On the other hand, $A^{(1)}_{i}$ contributes to first order in $\hbar$ and can emerge from the low energy limit of a more complete model (for example, in graphene, this is obtained from the low-energy limit of the tight-binding model and are responsible for pseudo-magnetic fields \cite{OLIVALEYVA20152645}).

\section{Semiclassical limit \label{sec:semiclass}}
To derive the ``ray approximation'' of electron optics to Eq. \eqref{eq:dirac_iso}, we proceed as usual by evaluating the asymptotic limit when $\hbar \rightarrow 0$. For this purpose, the semiclassical ansatz
\begin{eqnarray}
\label{eq:ansatz}
\psi(t,\boldsymbol{r}) = e^{{\tt i}\frac{S^{\pm}(t,\boldsymbol{r})}{\hbar}} \sum_{n=0}^{\infty} \hbar^{n} u_{n}(t,\boldsymbol{r}),
\end{eqnarray}
is inserted into Eq. \eqref{eq:dirac_iso},
where the amplitude $u$ is a bi-spinor and $S$ is the (real) phase \cite{maslov2001semi}. This approximation is valid when the electron wavelength is much smaller than typical length scales of the system. To leading order in $\hbar$ and transforming back to Cartesian coordinates, this yields the eikonal (or Hamilton-Jacobi) equation (see Appendix \ref{app:eom} for details):
\begin{align}
\label{eq:eikonal}
\partial_{t}S^{\pm}(t,\boldsymbol{x}) + \mathcal{H}^{\pm}(t,\boldsymbol{x}, \nabla_{\boldsymbol{x}} S) = 0 ,
\end{align}
where we defined the classical Hamiltonian
\begin{align}
\mathcal{H}^{\pm}(t,\boldsymbol{x}, \boldsymbol{p}) = \pm v_{D}\sqrt{ \frac{\boldsymbol{\pi}^{2}(\boldsymbol{x})}{\rho(\boldsymbol{x})} + m_{D}^{2}(\boldsymbol{x})v_{D}^{2} } + V(\boldsymbol{x}),
\end{align}
and the kinematic momentum as $\boldsymbol{\pi}(t,\boldsymbol{x}) = \boldsymbol{p} - \boldsymbol{A}^{(0)}(\boldsymbol{x})$. The superscript $\pm$ is the band index that stands for positive or negative energy bands (in the following, we consider positive energy solutions only). The Hamiltonian resembles that of a relativistic particle immersed in an electromagnetic field, except for the presence of $\rho$ that scales the momentum.

%\subsection{Classical-like equations of motion} 
Particle-like trajectories can be obtained from Eq. \eqref{eq:eikonal} via the method of characteristics. These equations are important physically because trajectories are orthogonal to wavefronts of the wave function \cite{doi:10.1119/10.0000781} and thus, are directly related to wave propagation. Letting $\boldsymbol{p} = \nabla_{\boldsymbol{x}} S$, the equations of motion are written as
\begin{eqnarray}
\dot{\boldsymbol{x}} := \frac{d \boldsymbol{x}}{dt}  = \nabla_{\boldsymbol{p}} \mathcal{H} \;,\;
\dot{\boldsymbol{p}} := \frac{d \boldsymbol{p}}{dt} = - \nabla_{\boldsymbol{x}} \mathcal{H}.
\end{eqnarray} 
After some manipulations given in Appendix \ref{app:eom}, they are given explicitly by
\begin{align}
\label{eq:eqm_x}
\frac{d \boldsymbol{x}}{dt} &= \frac{v_{D}}{\rho(\boldsymbol{x})}   \frac{\boldsymbol{\pi}(t,\boldsymbol{x})}{\sqrt{ \frac{\boldsymbol{\pi}^{2}(\boldsymbol{x})}{\rho(\boldsymbol{x})} + m_{D}^{2}(\boldsymbol{x})v_{D}^{2} }}, \\
\label{eq:lorentz}
\frac{d \boldsymbol{\pi}}{dt} 
&= \boldsymbol{E}(\boldsymbol{x}) + \dot{\boldsymbol{x}} \times \boldsymbol{B}(\boldsymbol{x}) + \boldsymbol{F}_{\mathrm{geom}}  + \boldsymbol{F}_{\mathrm{mass}},
\end{align}
where $\boldsymbol{E},\boldsymbol{B}$ are the electric and magnetic fields, respectively.
Eq. \eqref{eq:lorentz} corresponds to the relativistic Lorentz force equation for a particle immersed in an electromagnetic field with added fictitious forces due to geometry and to the space variation of the mass:
\begin{align}
\label{eq:Fgeom}
\boldsymbol{F}_{\mathrm{geom}} &= v_{D}   \frac{\boldsymbol{\pi}^{2}(\boldsymbol{x})}{2\sqrt{ \frac{\boldsymbol{\pi}^{2}(\boldsymbol{x})}{\rho(\boldsymbol{x})} + m_{D}^{2}(\boldsymbol{x})v_{D}^{2} }}  \frac{\nabla_{\boldsymbol{x}}\rho(\boldsymbol{x})}{\rho^{2}(\boldsymbol{x})}, \\
\label{eq:Fmass}
\boldsymbol{F}_{\mathrm{mass}} &= - \frac{v_{D}^{3}}{2\sqrt{ \frac{\boldsymbol{\pi}^{2}(\boldsymbol{x})}{\rho(\boldsymbol{x})} + m_{D}^{2}(\boldsymbol{x})v_{D}^{2} }} \nabla_{\boldsymbol{x}} m_{D}^{2}(\boldsymbol{x}).
\end{align}
Equation \eqref{eq:Fgeom} is an important result of this article. It implies that the leading order semiclassical approximation yields a theoretical description in terms of classical trajectories characterized by the Lorentz force equation \eqref{eq:lorentz}, with additional force terms and a space-dependent particle velocity. These features of the classical-like trajectories are not present for relativistic particles in flat space and thus, are direct consequences of the space curvature. They appear in equations via the presence of the isothermal coordinates metric component $\rho$. 

%\subsection{Graded refractive index}
Using Eq. \eqref{eq:eqm_x}, it is possible to evaluate the speed $v$ of the particle and define a graded index of refraction $n(\boldsymbol{x})$ as $v(\boldsymbol{x})=|\dot{\boldsymbol{x}}| = \frac{v_{D}}{n(\boldsymbol{x})}$. From this procedure, one finds a graded index of refraction given by
\begin{align}
\label{eq:ref_index}
n(\boldsymbol{x}) = \sqrt{\rho(\boldsymbol{x})} \sqrt{1+\frac{m_{D}^{2}(\boldsymbol{x})v_{D}^{2} \rho(\boldsymbol{x})}{\boldsymbol{\pi}^{2}(\boldsymbol{x})}}.
\end{align} 
In other words, the semiclassical propagation of fermions in Dirac matter is influenced by the geometry ($\rho$), the presence of the mass variation ($m_{D}$) and the momentum of the particle ($\boldsymbol{\pi}$).

%\subsection{Various limit of the semiclassical approximation}
It is instructive to look at certain limits of the formalism developed thus far. 
\begin{itemize}
	\item Massless limit:
	When there is no mass gap ($m_{D} = 0$), only the geometric force remains and becomes
	\begin{align}
	\boldsymbol{F}_{\mathrm{geom}} &= v_{D}   \frac{|\boldsymbol{\pi}(\boldsymbol{x})|}{2  }  \frac{\nabla_{\boldsymbol{x}}\rho(\boldsymbol{x})}{\rho^{\frac{3}{2}}(\boldsymbol{x})}.
	\end{align}
	Even more interesting is the fact that the index of refraction is directly related to the metric component as $n(\boldsymbol{x}) = \sqrt{\rho(\boldsymbol{x})}$. In other words, isothermal coordinates can be directly interpreted as the space dependence of the Dirac velocity. This also implies that in this limit, the particle velocity is solely dictated by the surface curvature. Other parameters of the model may influence its direction, but not how fast it propagates.
	\item Flat space limit:
	When the space is flat, the metric in isothermal coordinate is trivially given by $\rho = 1$. In this limit, $\boldsymbol{F}_{\mathrm{geom}} = 0$ and one recovers equations of motion given in Refs. \cite{PhysRevB.77.245413,REIJNDERS201865}.  
\end{itemize} 

\section{Numerical example: Gaussian out-of-plane deformation \label{sec:ex}}

To illustrate the behavior of trajectories, we consider a simple example where fermions are massless ($m_{D} = 0$) and where there are no external vector fields ($\boldsymbol{A}^{(0)} = 0$). In this case, the equations of motion can be simplified further by expressing them in a Newton-like form. By taking the time-derivative of the velocity \eqref{eq:eqm_x} and by changing the time variable from $t$ to $a$ to replicate the convention of Ref. \cite{doi:10.1119/1.14861} that $|d\boldsymbol{x}/da| = n(\boldsymbol{x})$, the equation of motion becomes
\begin{align}
\label{eq:newton}
\frac{d^{2} \boldsymbol{x}}{d a} = \nabla_{\boldsymbol{x}} \left[\frac{\rho(\boldsymbol{x})}{2} \right].
\end{align}
This equation is the same as the equation of motion for a light-ray in a graded refractive index medium, formulated in a Newton-like fashion with an effective potential given by $V_{\mathrm{eff}}(\boldsymbol{x}) = -\frac{n^{2}(\boldsymbol{x})}{2}$  \cite{doi:10.1119/1.14861}. Therefore, Eq. \eqref{eq:newton} connects electron optics in graphene to light optics in graded refractive index. As a consequence, the behavior of ``electronic-rays'' is analogous to that of light-rays because they obey the same equation to leading order in the semiclassical approximation. In Dirac matter, the index of refraction can be obtained by solving Eq. \eqref{eq:bel_eq} and by using Eq. \eqref{eq:rho}. In general, this task can be performed numerically. Least-square finite-element \cite{PhysRevE.103.013312} and finite-volume \cite{lorin2021} methods have been considered for that purpose.

\begin{figure}%[htb!]
	\begin{center}
		\includegraphics[width=0.5\textwidth]{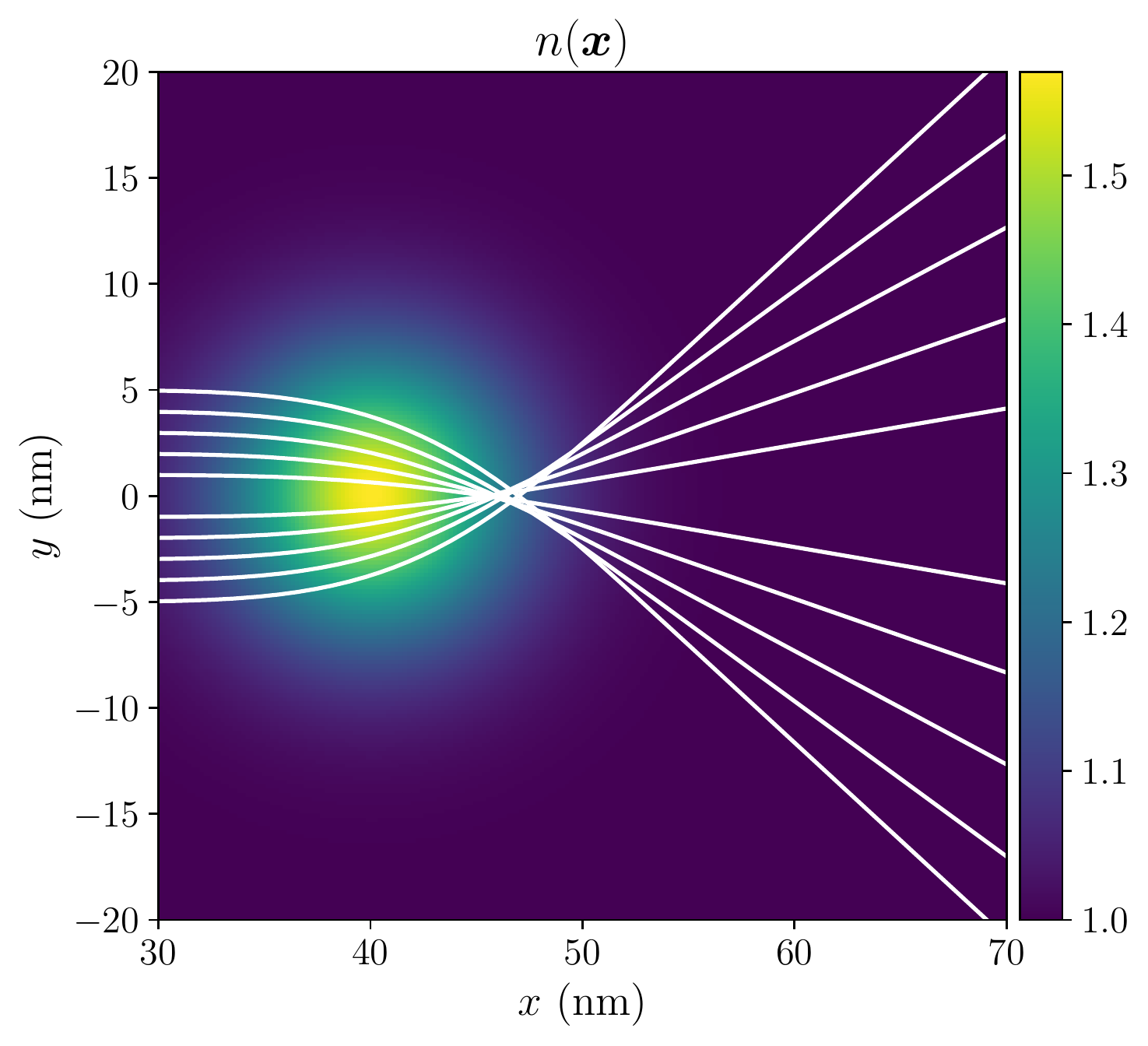}
	\end{center}
	\caption{Effective index of refraction corresponding to the deformation given in Eq. \eqref{eq:gaussian_def}, along with semiclassical trajectories (white lines).}
	\label{fig:index_ref}
\end{figure}

Next, we choose a particular configuration in which an incoming wave packet scatters off a Gaussian out-of-plane deformation with $X(\boldsymbol{x}) = Y(\boldsymbol{x}) = 0$ and
\begin{align}
\label{eq:gaussian_def}
Z(\boldsymbol{x}) &= C \exp \left(-\frac{|\boldsymbol{x} - \boldsymbol{x}_{p}|^{2}}{\alpha^{2}} \right),
\end{align}
where the amplitude of the deformation is set to $C = 10$ nm and the size is $\alpha = 10$ nm. The deformation is centered on the point $\boldsymbol{x}_{p}=(40 \; \mbox{nm}, 0)$. The Beltrami equation \ref{eq:bel_eq} is solved using the least-square finite-element method described in \cite{PhysRevE.103.013312} on a domain with a size of $L_{x} = L_{y} = 100$ nm using standard Lagrange order 2 elements (P$_{2}$-elements). To reach convergence, a grid of $1024 \times 1024$ points is considered. Initial conditions on the trajectories are chosen to represent an incoming wave packet with momentum in the $+x$-direction. They are given by $d \boldsymbol{x}(a)/da|_{a=0} = n(\boldsymbol{x}_{0}) \hat{\boldsymbol{x}}$ and $\boldsymbol{x}(0) = \boldsymbol{x}_{0} = (0,y_{n})$, where $y_{n} = n \delta y$ and $n \in \mathbb{Z}$. The numerical results for the corresponding index of refraction $n(\boldsymbol{x})$ are displayed in Fig. \ref{fig:index_ref}. In this figure, we also include semiclassical trajectories obtained by solving Eq. \eqref{eq:newton} numerically (using the LSODE package \cite{osti_15013302}). 

The numerical results clearly display the focusing of the wave packet as it scatters on the deformation and show the effect of the geometric force $\boldsymbol{F}_{\mathrm{geom}}$. This kind of behavior was expected from previous numerical studies, where the Dirac equation in curved space was solved numerically using an accurate pseudospectral method \cite{ANTOINE2020109412,PhysRevE.103.013312}. Therefore, the chosen deformation controls electronic rays like a lens. Physically, this can be understood by the fact that rays closer to the optical axis have to travel a longer effective distance owing to the geometry induced by the deformation, which in turn, lengthen their optical path. This modifies the wave-front and focuses the wave packet. This phenomenon is quite analogous to gravitational lensing around massive objects, where the gravitational field is responsible for modifying the space geometry. As a matter of fact, there exists theoretical approaches that describe gravitational lensing by introducing an effective graded index of refraction \cite{gravlens2008}, like we have done here in this work. 

In Fig. \ref{fig:index_ref}, one can also notice that semiclassical trajectories do not cross at the same point. Rather, the ones farther from the center are focused further on the optical axis. This is a signature of negative spherical aberrations induced by the spherical shape and the profile of the deformation. These aberrations may be a problem for some applications like the fermion microscope mentioned earlier, where an aberration-free system is desirable. Using our semiclassical approach, it is possible to inverse-engineer a specific deformation ($X,Y$ and $Z$) to eliminate these issues, using a metaheuristic algorithm (this will be presented in future work \cite{lorin2021}). More generally, we conjecture that the semiclassical approximation and the link with a refractive index could be used to design refractive optical-like elements such as lenses and wave guides to control electrons in 2D Dirac materials. Either the geometry (via $\rho$) or the external electromagnetic field can be used as control parameters to reach a desired graded refractive index. For even more flexibility and control, more complex (periodic) configurations similar to photonic crystals \cite{sakoda2004optical} could also be considered.  

Finally, when the trajectories are crossing, the semiclassical approximation breaks down. These are the well-known caustic points, where the asymptotic expansion in $\hbar$ does not converge \cite{maslov2001semi}. This can be circumvented by adapting the technique described in Ref. \cite{REIJNDERS201865}, by using the Gaussian beam method \cite{huang2012gaussian} or the frozen Gaussian approximation \cite{doi:10.1137/18M1222831}. The generalization of these methods to curved space is now under consideration.

\section{Conclusion \label{sec:conclu}}
To summarize, in this article, the semiclassical approximation was applied to 2D Dirac materials described by the Dirac equation in curved space-time. It was demonstrated that introducing isothermal coordinates simplifies the analysis significantly. In addition, the metric components in this coordinate system can be directly linked to an effective graded refractive index, leading to generalized electron optics. The transformation from Cartesian to isothermal coordinates can always be performed in 2D via quasi-conformal transformations. The latter can be evaluated numerically by solving the Beltrami equation. Compared to the case in flat space, a fictious geometric force appears in the classical equations of motion. This force can have a significant effect on electron trajectories, as was demonstrated in a specific example. 

In the future, the technique developed in this article will be used to design structures aiming at the control of electrons in Dirac matter. Indeed, evaluating the refractive index given in Eq. \eqref{eq:ref_index} is more efficient and sometimes, lead to more insights into the electron's behavior than a full numerical solution of the Dirac equation in curved space, as long as the conditions for the semiclassical approximation are fulfilled. Therefore, our work may have implications for important applications in nanoelectronics.

\begin{acknowledgments}
	The authors would like to thank Pierre L\'{e}vesque for numerous discussions on Dirac materials. This research was enabled in part by support provided by Compute Canada (\url{www.computecanada.ca}).
\end{acknowledgments}

\appendix

\section{Derivation of the Schr\"{o}dinger-like form of the Dirac equation \label{app:sch}}

In this section, we start from the covariant Dirac equation and derive its Schr\"{o}dinger-like form in isothermal coordinates. The 2D Dirac equation is given by 
\begin{align}
\label{eq:dirac_cov_app}
\left[ \texttt{i}v_{D} \hbar \bar{\gamma}^{\mu}(q)D_{\mu} - m_{D}(q)v_{D}^{2} \right]\psi(q) = 0,
\end{align} 
where $D_{\mu}=\partial_{\mu} + \Omega_{\mu} - {\tt i} A_{\mu}$ is the curved-space covariant derivative. The covariant derivative is a function of 
\begin{align}
\label{eq:spin_connection}
\Omega_{\mu}(q) = -\frac{{\tt i}}{4} \omega_{\mu}^{\ AB}(q) \sigma_{AB},
\end{align}
where $\sigma_{AB} = {\tt i}[\gamma_{A},\gamma_{B}]/2$ is the commutator of flat-space gamma-matrices.  The spin connection is defined as
\begin{align}
\omega_{\mu}^{\ AB}(q) =e_{\nu}^{A}(q) \left[   
\partial_{\mu}e^{\nu B}(q)  
+ \Gamma^{\nu}_{\ \mu \sigma}(q) e^{\sigma B}(q)
\right] \, ,
\end{align}
where we introduced the  Christoffel symbols 
\begin{align}
\label{eq:christo}
\Gamma^{\nu}_{\ \mu \sigma}(q) = \frac{g^{\nu \rho}(q)}{2} \left[
\partial_{\sigma}g_{\rho \mu}(q)
+ \partial_{\mu} g_{\rho \sigma}(q)
- \partial_{\rho}g_{\mu \sigma}(q)
\right] \, ,
\end{align}
In isothermal coordinates $r = (t,\boldsymbol{r})=(t,r_{1},r_{2})$, the metric is diagonal and can be written as a matrix:
\begin{align}
g(\boldsymbol{r}) = 
\begin{bmatrix}
1 & 0 &0 \\
0 & -\rho(\boldsymbol{r}) &0 \\
0 & 0 & -\rho(\boldsymbol{r})
\end{bmatrix}.
\end{align}
Using the vielbein relation to the metric $e_{\mu}^{A}(q) e^{B}_{\nu}(q) \eta_{AB} = g_{\mu \nu}(q)$, the natural vielbein can be easily found:
\begin{align}
e(\boldsymbol{r}) =
\begin{bmatrix}
1 & 0 & 0 \\
0 & \sqrt{\rho(\boldsymbol{r})} & 0 \\
0 & 0 &  \sqrt{\rho(\boldsymbol{r})}
\end{bmatrix} .
\end{align}
Using the fact that both the metric and vielbein are diagonal, and expanding the covariant indices, the Dirac equation becomes
\begin{align}
\label{eq:dirac_iso_app}
{\tt i}\hbar\partial_t \psi(t,{\boldsymbol r}) &= 
\biggl\{
-{\tt i}\hbar \frac{v_{D}}{\sqrt{\rho(\boldsymbol{r})}}\alpha^{i}   \left[\partial_{i} + \Omega_{i}({\boldsymbol r}) - {\tt i}  A_{i}({\boldsymbol r})\right] \nonumber \\
&
+ m_{D}(\boldsymbol{r})v_{D}^{2}\beta +   V({\boldsymbol r})  \mathbb{I}_{2}
\biggr\} \psi(t,{\boldsymbol r}) \, ,
\end{align}
where the affine spin connection components are
\begin{align}\label{Omegai}
\Omega_{1}(\boldsymbol{r}) = \frac{{\tt i}}{4} \frac{\partial_{2} \rho(\boldsymbol{r})}{|\rho(\boldsymbol{r})|} \beta \;\; , \;\;
\Omega_{2}(\boldsymbol{r}) = -\frac{{\tt i}}{4} \frac{\partial_{1} \rho(\boldsymbol{r})}{|\rho(\boldsymbol{r})|} \beta.
\end{align}
Thus, the Hamiltonian operator reads 
\begin{align}
\hat{H}_{g} &= 
-{\tt i}\hbar \frac{v_{D}}{\sqrt{\rho(\boldsymbol{r})}}\alpha^{i}   \left[\partial_{i} + \Omega_{i}({\boldsymbol r}) - {\tt i}  A_{i}({\boldsymbol r})\right] \nonumber \\
&+ m_{D}(\boldsymbol{r})v_{D}^{2}\beta +   V({\boldsymbol r}) \mathbb{I}_{2}.
\end{align}
However, this Hamiltonian is not Hermitian ($H_{g}^{\dagger} \neq H_{g}$). Rather, it is self-adjoint ($\langle \phi |\hat{H}_{g}^{\dagger} | \psi \rangle_{g} = \langle \phi |\hat{H}_{g} | \psi \rangle_{g}$) with respect to the $g$-scalar product \cite{PhysRevD.22.1922}
\begin{align}
\langle \phi |\psi \rangle_{g} &= \int dx \sqrt{\mathrm{det}(g)} \phi^{\dagger}(x) \psi(x).
\end{align}
Among other things, this implies that the $g$-norm $\Vert \psi \Vert_{g} = \langle \psi |\psi \rangle_{g}$ is conserved by the time-evolution given by the Dirac equation in curved space \cite{PhysRevD.22.1922}. 

To make a connection with the low-energy limit of Dirac matter tight-binding models or to develop numerical schemes that preserves unitarity, it is convenient to transform the Hamiltonian in such a way that the $L_{2}$ scalar product is preserved. This can be performed because the Hamiltonian is a pseudo-Hermitian operator \cite{PhysRevD.83.105002}. In this case, the Hamiltonian and the wave function can be transformed as
\begin{align}
\hat{H} &= \eta \hat{H}_{g} \eta^{-1}, \\
\psi_{\eta}(t,{\boldsymbol r}) &= \eta \psi(t,{\boldsymbol r}),
\end{align} 
such that the $g$- and $L_{2}$- scalar products are identical:
\begin{align}
\langle \phi | \psi \rangle_{g} = \langle \phi_{\eta} | \psi_{\eta} \rangle_{L^{2}}.
\end{align}
In this new representation, the Hamiltonian $\hat{H}$ is self-adjoint with respect to the $L_{2}$-product, as in usual quantum mechanics. The transformation that allows for this property is $\eta = (\mathrm{det}g)^{\frac{1}{4}}$ \cite{PhysRevD.83.105002}. This gives
\begin{align}
\hat{H} &= 
-{\tt i}\hbar \frac{v_{D}}{\sqrt{\rho(\boldsymbol{r})}}\alpha^{1}   \left[\partial_{1} - \frac{\partial_{1} \sqrt{\mathrm{det}g}}{2 \sqrt{\mathrm{det}g} } + \Omega_{1}({\boldsymbol r}) - {\tt i}  A_{1}({\boldsymbol r})\right] \nonumber \\
&\;\;\;-{\tt i}\hbar \frac{v_{D}}{\sqrt{\rho(\boldsymbol{r})}}\alpha^{2}   \left[\partial_{i} - \frac{\partial_{2} \sqrt{\mathrm{det}g}}{2 \sqrt{\mathrm{det}g} } + \Omega_{2}({\boldsymbol r}) - {\tt i}  A_{2}({\boldsymbol r})\right] \nonumber \\
&+ m_{D}(\boldsymbol{r})v_{D}^{2}\beta +   V({\boldsymbol r}) \mathbb{I}_{2}.
\end{align}
Finally, using the facts that $\mathrm{det}(g) = \rho$ and $\alpha_{i}\beta = {\tt i}\epsilon_{i3k} \sigma_{k}$, we get
\begin{align}
\hat{H} &= 
-{\tt i}\hbar \frac{v_{D}}{\sqrt{\rho(\boldsymbol{r})}}\alpha^{i}   \left[\partial_{i} -\frac{1}{4} \partial_{i} \ln \big(\rho(\boldsymbol{r}) \bigr) - {\tt i}  A_{i}({\boldsymbol r})\right] \nonumber \\
&+ m_{D}(\boldsymbol{r})v_{D}^{2}\beta +  V({\boldsymbol r}) \mathbb{I}_{2} .
\end{align}

\section{Derivation of the equation of motion \label{app:eom}}

The starting point of the derivation is the Dirac equation in Schr\"{o}dinger-like form:
\begin{align}
\label{eq:dirac_sa}
{\tt i} \hbar \partial_{t} \psi(t,\boldsymbol{r}) &= \nonumber \\
 &\bigg\{-{\tt i}\hbar \frac{v_{D}}{\sqrt{\rho(\boldsymbol{r})}}\alpha^{i}   \left[\partial_{i} -\frac{1}{4} \partial_{i} \ln \big(\rho(\boldsymbol{r}) \bigr) \right] \nonumber \\
 &-\frac{v_{D}}{\sqrt{\rho(\boldsymbol{r})}}\alpha^{i}   \left[  A^{(0)}_{i}({\boldsymbol r}) +   \hbar  A^{(1)}_{i}({\boldsymbol r})\right] \nonumber \\
&+ m_{D}(\boldsymbol{r})v_{D}^{2}\beta  +    V({\boldsymbol r}) \mathbb{I}_{2}\biggr\} \psi(t,\boldsymbol{r}).
\end{align}
Inserting the ansatz 
\begin{eqnarray}
\label{eq:ansatz_app}
\psi(t,\boldsymbol{r}) = e^{{\tt i}\frac{S^{\pm}(t,\boldsymbol{r})}{\hbar}} \sum_{n=0}^{\infty} \hbar^{n} u_{n}(t,\boldsymbol{r}),
\end{eqnarray}
into Eq. \eqref{eq:dirac_sa}, we get to $O(\hbar^{0})$ the following system of equations:
\begin{align}
\label{eq:real}
\mathcal{G}(\boldsymbol{r}) u_{0}(t,\boldsymbol{r}) =0 ,
\end{align}
where
\begin{align}
\mathcal{G}(\boldsymbol{r}) &:= \partial_{t}S(t,\boldsymbol{r}) +  \frac{v_{D} \alpha^{i}}{\sqrt{\rho(\boldsymbol{r})}}   \left[ \partial_{i} S(t,\boldsymbol{r}) - A_{i}^{(0)}(\boldsymbol{r}) \right]  \nonumber \\
& + m_{D}(\boldsymbol{r})v_{D}^{2}\beta + V(\boldsymbol{r}).
\end{align}
For completion, we also display the contribution at $O(\hbar^{1})$:
\begin{align}
{\tt i}\left[\partial_{t} + \frac{v_{D}\alpha^{i}}{\sqrt{\rho(\boldsymbol{r})}} \left(\partial_{i} + \tilde{\Omega}_{i}(\boldsymbol{r}) 
- {\tt i}A_{i}^{(1)}(\boldsymbol{r}) \right)\right] u_{0} \nonumber \\
- \mathcal{G}(\boldsymbol{r}) u_{1} = 0.
\end{align}
A dynamical equation for $S$ can be obtained by noticing that Eq. \eqref{eq:real} is a homogeneous system of linear equations ($\mathcal{G}$ is a two-by-two matrix in spinor space), having a non-trivial solution only if the determinant in spinor space is zero \cite{huang2012gaussian}. Evaluating the determinant $\det \mathcal{G} = 0$ gives
\begin{align}
\label{eq:eikonal_iso}
\partial_{t}S^{\pm}(t,\boldsymbol{r}) = \mathcal{H}^{\pm}(t,\boldsymbol{r}, \nabla_{\boldsymbol{r}} S^{\pm}) ,
\end{align}
where we defined the classical Hamiltonian
\begin{align}
\mathcal{H}^{\pm}(t,\boldsymbol{q}, \boldsymbol{p}) = \pm v_{D}\sqrt{ \frac{\boldsymbol{\pi}^{2}(\boldsymbol{q})}{\rho(\boldsymbol{q})} + m_{D}^{2}(\boldsymbol{q})v_{D}^{2} } + V(\boldsymbol{q}),
\end{align}
and the kinematic momentum as $\boldsymbol{\pi}(t,\boldsymbol{q}) = \boldsymbol{p} - \boldsymbol{A}(\boldsymbol{q})$. Then, using the fact that all the terms in this equation are scalars, it can be expressed in Cartesian coordinates in a straightforward way by performing a change of variable $\boldsymbol{r} \rightarrow \boldsymbol{x}$. We get the eikonal equation in Cartesian coordinates:
\begin{align}
\label{eq:eikonal_app}
\partial_{t}S^{\pm}(t,\boldsymbol{x}) + \mathcal{H}^{\pm}(t,\boldsymbol{x}, \nabla_{\boldsymbol{x}} S^{\pm}) = 0 .
\end{align}
This corresponds to the Hamilton-Jacobi equation for a relativistic particle in an electromagnetic field, with an added function $\rho$ that scales the momentum.

Particle-like trajectories can be obtained from this equation via the method of characteristics. Letting $\boldsymbol{p} = \nabla_{\boldsymbol{x}} S$, the equations of motion are written as
\begin{eqnarray}
\dot{\boldsymbol{x}} := \frac{d \boldsymbol{x}}{dt}  = \nabla_{\boldsymbol{p}} \mathcal{H} \;,\;
\dot{\boldsymbol{p}} := \frac{d \boldsymbol{p}}{dt} = - \nabla_{\boldsymbol{x}} \mathcal{H}.
\end{eqnarray} 
Defining the kinematic momentum as $\boldsymbol{\pi}(t,\boldsymbol{x}) = \boldsymbol{p} - \boldsymbol{A}(\boldsymbol{x})$,  they are given explicitly by
\begin{align}
\label{eq:eqm_x_app}
\frac{d \boldsymbol{x}}{dt} &= \frac{v_{D}}{\rho(\boldsymbol{x})}   \frac{\boldsymbol{\pi}(t,\boldsymbol{x})}{\sqrt{ \frac{\boldsymbol{\pi}^{2}(\boldsymbol{x})}{\rho(\boldsymbol{x})} + m_{D}^{2}(\boldsymbol{x})v_{D}^{2} }}, \\
\label{eq:eqm_p_app}
\frac{d \boldsymbol{p}}{dt} &= -\nabla_{\boldsymbol{x}}V + \nabla_{\boldsymbol{x},\boldsymbol{A}}(\dot{\boldsymbol{x}} \cdot \boldsymbol{A}) 
+ \frac{\dot{\boldsymbol{x}}\cdot \boldsymbol{\pi}(\boldsymbol{x})}{2} \frac{\nabla_{\boldsymbol{x}}\rho(\boldsymbol{x})}{\rho(\boldsymbol{x})} , \nonumber \\
& - \frac{v_{D}^{3}}{2\sqrt{ \frac{\boldsymbol{\pi}^{2}(\boldsymbol{x})}{\rho(\boldsymbol{x})} + m_{D}^{2}(\boldsymbol{x})v_{D}^{2} }} \nabla_{\boldsymbol{x}} m_{D}^{2}(\boldsymbol{x}),
\end{align}
where $\nabla_{\boldsymbol{x},\boldsymbol{A}}$ is Feynman's notation, meaning that the gradient is applied on the vector field $\boldsymbol{A}$. The force equation can be obtained by evaluating the time derivative of the kinematic momentum:
\begin{align}
\frac{d \boldsymbol{\pi}}{dt} &= \dot{\boldsymbol{p}} - \frac{d\boldsymbol{A}(\boldsymbol{x})}{dt}, \\
&= \dot{\boldsymbol{p}} - (\dot{\boldsymbol{x}} \cdot \nabla) \boldsymbol{A}(\boldsymbol{x}), \\
\label{eq:lorentz_app}
&= \boldsymbol{E}(\boldsymbol{x}) + \dot{\boldsymbol{x}} \times \boldsymbol{B}(\boldsymbol{x}) + \boldsymbol{F}_{\mathrm{geom}}  + \boldsymbol{F}_{\mathrm{mass}},
\end{align}
where the vectorial identity $\nabla_{\boldsymbol{W}}(\boldsymbol{V} \boldsymbol{W}) = \boldsymbol{V} \times \nabla \times \boldsymbol{W} + (\boldsymbol{V} \cdot \nabla) \boldsymbol{W} $ has been used to get the last equation. Eq. \eqref{eq:lorentz_app} corresponds to the relativistic Lorentz force equation for a particle immersed in an electromagnetic field with added fictitious forces due to geometry and to the space variation of the mass:
\begin{align}
\boldsymbol{F}_{\mathrm{geom}} &= v_{D}   \frac{\boldsymbol{\pi}^{2}(\boldsymbol{x})}{2\sqrt{ \frac{\boldsymbol{\pi}^{2}(\boldsymbol{x})}{\rho(\boldsymbol{x})} + m_{D}^{2}(\boldsymbol{x})v_{D}^{2} }}  \frac{\nabla_{\boldsymbol{x}}\rho(\boldsymbol{x})}{\rho^{2}(\boldsymbol{x})}, \\
\boldsymbol{F}_{\mathrm{mass}} &= - \frac{v_{D}^{3}}{2\sqrt{ \frac{\boldsymbol{\pi}^{2}(\boldsymbol{x})}{\rho(\boldsymbol{x})} + m_{D}^{2}(\boldsymbol{x})v_{D}^{2} }} \nabla_{\boldsymbol{x}} m_{D}^{2}(\boldsymbol{x}).
\end{align}

\bibliography{refs}

\begin{thebibliography}{66}%
\makeatletter
\providecommand \@ifxundefined [1]{%
 \@ifx{#1\undefined}
}%
\providecommand \@ifnum [1]{%
 \ifnum #1\expandafter \@firstoftwo
 \else \expandafter \@secondoftwo
 \fi
}%
\providecommand \@ifx [1]{%
 \ifx #1\expandafter \@firstoftwo
 \else \expandafter \@secondoftwo
 \fi
}%
\providecommand \natexlab [1]{#1}%
\providecommand \enquote  [1]{``#1''}%
\providecommand \bibnamefont  [1]{#1}%
\providecommand \bibfnamefont [1]{#1}%
\providecommand \citenamefont [1]{#1}%
\providecommand \href@noop [0]{\@secondoftwo}%
\providecommand \href [0]{\begingroup \@sanitize@url \@href}%
\providecommand \@href[1]{\@@startlink{#1}\@@href}%
\providecommand \@@href[1]{\endgroup#1\@@endlink}%
\providecommand \@sanitize@url [0]{\catcode `\\12\catcode `\$12\catcode
  `\&12\catcode `\#12\catcode `\^12\catcode `\_12\catcode `\%12\relax}%
\providecommand \@@startlink[1]{}%
\providecommand \@@endlink[0]{}%
\providecommand \url  [0]{\begingroup\@sanitize@url \@url }%
\providecommand \@url [1]{\endgroup\@href {#1}{\urlprefix }}%
\providecommand \urlprefix  [0]{URL }%
\providecommand \Eprint [0]{\href }%
\providecommand \doibase [0]{https://doi.org/}%
\providecommand \selectlanguage [0]{\@gobble}%
\providecommand \bibinfo  [0]{\@secondoftwo}%
\providecommand \bibfield  [0]{\@secondoftwo}%
\providecommand \translation [1]{[#1]}%
\providecommand \BibitemOpen [0]{}%
\providecommand \bibitemStop [0]{}%
\providecommand \bibitemNoStop [0]{.\EOS\space}%
\providecommand \EOS [0]{\spacefactor3000\relax}%
\providecommand \BibitemShut  [1]{\csname bibitem#1\endcsname}%
\let\auto@bib@innerbib\@empty
%</preamble>
\bibitem [{\citenamefont {Wehling}\ \emph {et~al.}(2014)\citenamefont
  {Wehling}, \citenamefont {Black-Schaffer},\ and\ \citenamefont
  {Balatsky}}]{doi:10.1080/00018732.2014.927109}%
  \BibitemOpen
  \bibfield  {author} {\bibinfo {author} {\bibfnamefont {T.}~\bibnamefont
  {Wehling}}, \bibinfo {author} {\bibfnamefont {A.}~\bibnamefont
  {Black-Schaffer}},\ and\ \bibinfo {author} {\bibfnamefont {A.}~\bibnamefont
  {Balatsky}},\ }\bibfield  {title} {\bibinfo {title} {Dirac materials},\
  }\href {https://doi.org/10.1080/00018732.2014.927109} {\bibfield  {journal}
  {\bibinfo  {journal} {Advances in Physics}\ }\textbf {\bibinfo {volume}
  {63}},\ \bibinfo {pages} {1} (\bibinfo {year} {2014})}\BibitemShut {NoStop}%
\bibitem [{\citenamefont {Cayssol}(2013)}]{CAYSSOL2013760}%
  \BibitemOpen
  \bibfield  {author} {\bibinfo {author} {\bibfnamefont {J.}~\bibnamefont
  {Cayssol}},\ }\bibfield  {title} {\bibinfo {title} {Introduction to dirac
  materials and topological insulators},\ }\href
  {https://doi.org/https://doi.org/10.1016/j.crhy.2013.09.012} {\bibfield
  {journal} {\bibinfo  {journal} {Comptes Rendus Physique}\ }\textbf {\bibinfo
  {volume} {14}},\ \bibinfo {pages} {760} (\bibinfo {year} {2013})},\ \bibinfo
  {note} {topological insulators / Isolants topologiques}\BibitemShut {NoStop}%
\bibitem [{\citenamefont {Wallace}(1947)}]{PhysRev.71.622}%
  \BibitemOpen
  \bibfield  {author} {\bibinfo {author} {\bibfnamefont {P.~R.}\ \bibnamefont
  {Wallace}},\ }\bibfield  {title} {\bibinfo {title} {The band theory of
  graphite},\ }\href {https://doi.org/10.1103/PhysRev.71.622} {\bibfield
  {journal} {\bibinfo  {journal} {Phys. Rev.}\ }\textbf {\bibinfo {volume}
  {71}},\ \bibinfo {pages} {622} (\bibinfo {year} {1947})}\BibitemShut
  {NoStop}%
\bibitem [{\citenamefont {Geim}\ and\ \citenamefont
  {Novoselov}()}]{doi:10.1142/9789814287005_0002}%
  \BibitemOpen
  \bibfield  {author} {\bibinfo {author} {\bibfnamefont {A.~K.}\ \bibnamefont
  {Geim}}\ and\ \bibinfo {author} {\bibfnamefont {K.~S.}\ \bibnamefont
  {Novoselov}},\ }\bibfield  {title} {\bibinfo {title} {The rise of graphene},\
  }in\ \href {https://doi.org/10.1142/9789814287005_0002} {\emph {\bibinfo
  {booktitle} {Nanoscience and Technology}}},\ pp.\ \bibinfo {pages}
  {11--19}\BibitemShut {NoStop}%
\bibitem [{\citenamefont {Castro~Neto}\ \emph {et~al.}(2009)\citenamefont
  {Castro~Neto}, \citenamefont {Guinea}, \citenamefont {Peres}, \citenamefont
  {Novoselov},\ and\ \citenamefont {Geim}}]{RevModPhys.81.109}%
  \BibitemOpen
  \bibfield  {author} {\bibinfo {author} {\bibfnamefont {A.~H.}\ \bibnamefont
  {Castro~Neto}}, \bibinfo {author} {\bibfnamefont {F.}~\bibnamefont {Guinea}},
  \bibinfo {author} {\bibfnamefont {N.~M.~R.}\ \bibnamefont {Peres}}, \bibinfo
  {author} {\bibfnamefont {K.~S.}\ \bibnamefont {Novoselov}},\ and\ \bibinfo
  {author} {\bibfnamefont {A.~K.}\ \bibnamefont {Geim}},\ }\bibfield  {title}
  {\bibinfo {title} {The electronic properties of graphene},\ }\href
  {https://doi.org/10.1103/RevModPhys.81.109} {\bibfield  {journal} {\bibinfo
  {journal} {Rev. Mod. Phys.}\ }\textbf {\bibinfo {volume} {81}},\ \bibinfo
  {pages} {109} (\bibinfo {year} {2009})}\BibitemShut {NoStop}%
\bibitem [{\citenamefont {Hasan}\ and\ \citenamefont
  {Kane}(2010)}]{RevModPhys.82.3045}%
  \BibitemOpen
  \bibfield  {author} {\bibinfo {author} {\bibfnamefont {M.~Z.}\ \bibnamefont
  {Hasan}}\ and\ \bibinfo {author} {\bibfnamefont {C.~L.}\ \bibnamefont
  {Kane}},\ }\bibfield  {title} {\bibinfo {title} {Colloquium: Topological
  insulators},\ }\href {https://doi.org/10.1103/RevModPhys.82.3045} {\bibfield
  {journal} {\bibinfo  {journal} {Rev. Mod. Phys.}\ }\textbf {\bibinfo {volume}
  {82}},\ \bibinfo {pages} {3045} (\bibinfo {year} {2010})}\BibitemShut
  {NoStop}%
\bibitem [{\citenamefont {Bansil}\ \emph {et~al.}(2016)\citenamefont {Bansil},
  \citenamefont {Lin},\ and\ \citenamefont {Das}}]{RevModPhys.88.021004}%
  \BibitemOpen
  \bibfield  {author} {\bibinfo {author} {\bibfnamefont {A.}~\bibnamefont
  {Bansil}}, \bibinfo {author} {\bibfnamefont {H.}~\bibnamefont {Lin}},\ and\
  \bibinfo {author} {\bibfnamefont {T.}~\bibnamefont {Das}},\ }\bibfield
  {title} {\bibinfo {title} {Colloquium: Topological band theory},\ }\href
  {https://doi.org/10.1103/RevModPhys.88.021004} {\bibfield  {journal}
  {\bibinfo  {journal} {Rev. Mod. Phys.}\ }\textbf {\bibinfo {volume} {88}},\
  \bibinfo {pages} {021004} (\bibinfo {year} {2016})}\BibitemShut {NoStop}%
\bibitem [{\citenamefont {Qi}\ and\ \citenamefont
  {Zhang}(2011)}]{RevModPhys.83.1057}%
  \BibitemOpen
  \bibfield  {author} {\bibinfo {author} {\bibfnamefont {X.-L.}\ \bibnamefont
  {Qi}}\ and\ \bibinfo {author} {\bibfnamefont {S.-C.}\ \bibnamefont {Zhang}},\
  }\bibfield  {title} {\bibinfo {title} {Topological insulators and
  superconductors},\ }\href {https://doi.org/10.1103/RevModPhys.83.1057}
  {\bibfield  {journal} {\bibinfo  {journal} {Rev. Mod. Phys.}\ }\textbf
  {\bibinfo {volume} {83}},\ \bibinfo {pages} {1057} (\bibinfo {year}
  {2011})}\BibitemShut {NoStop}%
\bibitem [{\citenamefont {Armitage}\ \emph {et~al.}(2018)\citenamefont
  {Armitage}, \citenamefont {Mele},\ and\ \citenamefont
  {Vishwanath}}]{RevModPhys.90.015001}%
  \BibitemOpen
  \bibfield  {author} {\bibinfo {author} {\bibfnamefont {N.~P.}\ \bibnamefont
  {Armitage}}, \bibinfo {author} {\bibfnamefont {E.~J.}\ \bibnamefont {Mele}},\
  and\ \bibinfo {author} {\bibfnamefont {A.}~\bibnamefont {Vishwanath}},\
  }\bibfield  {title} {\bibinfo {title} {Weyl and dirac semimetals in
  three-dimensional solids},\ }\href
  {https://doi.org/10.1103/RevModPhys.90.015001} {\bibfield  {journal}
  {\bibinfo  {journal} {Rev. Mod. Phys.}\ }\textbf {\bibinfo {volume} {90}},\
  \bibinfo {pages} {015001} (\bibinfo {year} {2018})}\BibitemShut {NoStop}%
\bibitem [{\citenamefont {Balatsky}\ \emph {et~al.}(2006)\citenamefont
  {Balatsky}, \citenamefont {Vekhter},\ and\ \citenamefont
  {Zhu}}]{RevModPhys.78.373}%
  \BibitemOpen
  \bibfield  {author} {\bibinfo {author} {\bibfnamefont {A.~V.}\ \bibnamefont
  {Balatsky}}, \bibinfo {author} {\bibfnamefont {I.}~\bibnamefont {Vekhter}},\
  and\ \bibinfo {author} {\bibfnamefont {J.-X.}\ \bibnamefont {Zhu}},\
  }\bibfield  {title} {\bibinfo {title} {Impurity-induced states in
  conventional and unconventional superconductors},\ }\href
  {https://doi.org/10.1103/RevModPhys.78.373} {\bibfield  {journal} {\bibinfo
  {journal} {Rev. Mod. Phys.}\ }\textbf {\bibinfo {volume} {78}},\ \bibinfo
  {pages} {373} (\bibinfo {year} {2006})}\BibitemShut {NoStop}%
\bibitem [{\citenamefont {Volovik}(1992)}]{volovik1992exotic}%
  \BibitemOpen
  \bibfield  {author} {\bibinfo {author} {\bibfnamefont {G.~E.}\ \bibnamefont
  {Volovik}},\ }\href@noop {} {\emph {\bibinfo {title} {Exotic properties of
  superfluid 3He}}},\ Vol.~\bibinfo {volume} {1}\ (\bibinfo  {publisher} {World
  Scientific},\ \bibinfo {year} {1992})\BibitemShut {NoStop}%
\bibitem [{\citenamefont {Uehlinger}\ \emph {et~al.}(2013)\citenamefont
  {Uehlinger}, \citenamefont {Jotzu}, \citenamefont {Messer}, \citenamefont
  {Greif}, \citenamefont {Hofstetter}, \citenamefont {Bissbort},\ and\
  \citenamefont {Esslinger}}]{PhysRevLett.111.185307}%
  \BibitemOpen
  \bibfield  {author} {\bibinfo {author} {\bibfnamefont {T.}~\bibnamefont
  {Uehlinger}}, \bibinfo {author} {\bibfnamefont {G.}~\bibnamefont {Jotzu}},
  \bibinfo {author} {\bibfnamefont {M.}~\bibnamefont {Messer}}, \bibinfo
  {author} {\bibfnamefont {D.}~\bibnamefont {Greif}}, \bibinfo {author}
  {\bibfnamefont {W.}~\bibnamefont {Hofstetter}}, \bibinfo {author}
  {\bibfnamefont {U.}~\bibnamefont {Bissbort}},\ and\ \bibinfo {author}
  {\bibfnamefont {T.}~\bibnamefont {Esslinger}},\ }\bibfield  {title} {\bibinfo
  {title} {Artificial graphene with tunable interactions},\ }\href
  {https://doi.org/10.1103/PhysRevLett.111.185307} {\bibfield  {journal}
  {\bibinfo  {journal} {Phys. Rev. Lett.}\ }\textbf {\bibinfo {volume} {111}},\
  \bibinfo {pages} {185307} (\bibinfo {year} {2013})}\BibitemShut {NoStop}%
\bibitem [{\citenamefont {Raghu}\ and\ \citenamefont
  {Haldane}(2008)}]{PhysRevA.78.033834}%
  \BibitemOpen
  \bibfield  {author} {\bibinfo {author} {\bibfnamefont {S.}~\bibnamefont
  {Raghu}}\ and\ \bibinfo {author} {\bibfnamefont {F.~D.~M.}\ \bibnamefont
  {Haldane}},\ }\bibfield  {title} {\bibinfo {title} {Analogs of
  quantum-hall-effect edge states in photonic crystals},\ }\href
  {https://doi.org/10.1103/PhysRevA.78.033834} {\bibfield  {journal} {\bibinfo
  {journal} {Phys. Rev. A}\ }\textbf {\bibinfo {volume} {78}},\ \bibinfo
  {pages} {033834} (\bibinfo {year} {2008})}\BibitemShut {NoStop}%
\bibitem [{\citenamefont {Sepkhanov}\ \emph {et~al.}(2007)\citenamefont
  {Sepkhanov}, \citenamefont {Bazaliy},\ and\ \citenamefont
  {Beenakker}}]{PhysRevA.75.063813}%
  \BibitemOpen
  \bibfield  {author} {\bibinfo {author} {\bibfnamefont {R.~A.}\ \bibnamefont
  {Sepkhanov}}, \bibinfo {author} {\bibfnamefont {Y.~B.}\ \bibnamefont
  {Bazaliy}},\ and\ \bibinfo {author} {\bibfnamefont {C.~W.~J.}\ \bibnamefont
  {Beenakker}},\ }\bibfield  {title} {\bibinfo {title} {Extremal transmission
  at the dirac point of a photonic band structure},\ }\href
  {https://doi.org/10.1103/PhysRevA.75.063813} {\bibfield  {journal} {\bibinfo
  {journal} {Phys. Rev. A}\ }\textbf {\bibinfo {volume} {75}},\ \bibinfo
  {pages} {063813} (\bibinfo {year} {2007})}\BibitemShut {NoStop}%
\bibitem [{\citenamefont {Zhang}(2008)}]{PhysRevLett.100.113903}%
  \BibitemOpen
  \bibfield  {author} {\bibinfo {author} {\bibfnamefont {X.}~\bibnamefont
  {Zhang}},\ }\bibfield  {title} {\bibinfo {title} {Observing zitterbewegung
  for photons near the dirac point of a two-dimensional photonic crystal},\
  }\href {https://doi.org/10.1103/PhysRevLett.100.113903} {\bibfield  {journal}
  {\bibinfo  {journal} {Phys. Rev. Lett.}\ }\textbf {\bibinfo {volume} {100}},\
  \bibinfo {pages} {113903} (\bibinfo {year} {2008})}\BibitemShut {NoStop}%
\bibitem [{\citenamefont {Rechtsman}\ \emph {et~al.}(2013)\citenamefont
  {Rechtsman}, \citenamefont {Plotnik}, \citenamefont {Zeuner}, \citenamefont
  {Song}, \citenamefont {Chen}, \citenamefont {Szameit},\ and\ \citenamefont
  {Segev}}]{PhysRevLett.111.103901}%
  \BibitemOpen
  \bibfield  {author} {\bibinfo {author} {\bibfnamefont {M.~C.}\ \bibnamefont
  {Rechtsman}}, \bibinfo {author} {\bibfnamefont {Y.}~\bibnamefont {Plotnik}},
  \bibinfo {author} {\bibfnamefont {J.~M.}\ \bibnamefont {Zeuner}}, \bibinfo
  {author} {\bibfnamefont {D.}~\bibnamefont {Song}}, \bibinfo {author}
  {\bibfnamefont {Z.}~\bibnamefont {Chen}}, \bibinfo {author} {\bibfnamefont
  {A.}~\bibnamefont {Szameit}},\ and\ \bibinfo {author} {\bibfnamefont
  {M.}~\bibnamefont {Segev}},\ }\bibfield  {title} {\bibinfo {title}
  {Topological creation and destruction of edge states in photonic graphene},\
  }\href {https://doi.org/10.1103/PhysRevLett.111.103901} {\bibfield  {journal}
  {\bibinfo  {journal} {Phys. Rev. Lett.}\ }\textbf {\bibinfo {volume} {111}},\
  \bibinfo {pages} {103901} (\bibinfo {year} {2013})}\BibitemShut {NoStop}%
\bibitem [{\citenamefont {Khanikaev}\ \emph {et~al.}(2013)\citenamefont
  {Khanikaev}, \citenamefont {Mousavi}, \citenamefont {Tse}, \citenamefont
  {Kargarian}, \citenamefont {MacDonald},\ and\ \citenamefont
  {Shvets}}]{khanikaev2013photonic}%
  \BibitemOpen
  \bibfield  {author} {\bibinfo {author} {\bibfnamefont {A.~B.}\ \bibnamefont
  {Khanikaev}}, \bibinfo {author} {\bibfnamefont {S.~H.}\ \bibnamefont
  {Mousavi}}, \bibinfo {author} {\bibfnamefont {W.-K.}\ \bibnamefont {Tse}},
  \bibinfo {author} {\bibfnamefont {M.}~\bibnamefont {Kargarian}}, \bibinfo
  {author} {\bibfnamefont {A.~H.}\ \bibnamefont {MacDonald}},\ and\ \bibinfo
  {author} {\bibfnamefont {G.}~\bibnamefont {Shvets}},\ }\bibfield  {title}
  {\bibinfo {title} {Photonic topological insulators},\ }\href
  {https://doi.org/https://doi.org/10.1038/nmat3520} {\bibfield  {journal}
  {\bibinfo  {journal} {Nature materials}\ }\textbf {\bibinfo {volume} {12}},\
  \bibinfo {pages} {233} (\bibinfo {year} {2013})}\BibitemShut {NoStop}%
\bibitem [{\citenamefont {Mousavi}\ \emph {et~al.}(2015)\citenamefont
  {Mousavi}, \citenamefont {Khanikaev},\ and\ \citenamefont
  {Wang}}]{mousavi2015topologically}%
  \BibitemOpen
  \bibfield  {author} {\bibinfo {author} {\bibfnamefont {S.~H.}\ \bibnamefont
  {Mousavi}}, \bibinfo {author} {\bibfnamefont {A.~B.}\ \bibnamefont
  {Khanikaev}},\ and\ \bibinfo {author} {\bibfnamefont {Z.}~\bibnamefont
  {Wang}},\ }\bibfield  {title} {\bibinfo {title} {Topologically protected
  elastic waves in phononic metamaterials},\ }\href
  {https://doi.org/https://doi.org/10.1038/ncomms9682} {\bibfield  {journal}
  {\bibinfo  {journal} {Nature communications}\ }\textbf {\bibinfo {volume}
  {6}},\ \bibinfo {pages} {1} (\bibinfo {year} {2015})}\BibitemShut {NoStop}%
\bibitem [{\citenamefont {Braun}\ \emph {et~al.}(1999)\citenamefont {Braun},
  \citenamefont {Su},\ and\ \citenamefont {Grobe}}]{PhysRevA.59.604}%
  \BibitemOpen
  \bibfield  {author} {\bibinfo {author} {\bibfnamefont {J.~W.}\ \bibnamefont
  {Braun}}, \bibinfo {author} {\bibfnamefont {Q.}~\bibnamefont {Su}},\ and\
  \bibinfo {author} {\bibfnamefont {R.}~\bibnamefont {Grobe}},\ }\bibfield
  {title} {\bibinfo {title} {Numerical approach to solve the time-dependent
  dirac equation},\ }\href {https://doi.org/10.1103/PhysRevA.59.604} {\bibfield
   {journal} {\bibinfo  {journal} {Phys. Rev. A}\ }\textbf {\bibinfo {volume}
  {59}},\ \bibinfo {pages} {604} (\bibinfo {year} {1999})}\BibitemShut
  {NoStop}%
\bibitem [{\citenamefont {Succi}\ and\ \citenamefont
  {Benzi}(1993)}]{SUCCI1993327}%
  \BibitemOpen
  \bibfield  {author} {\bibinfo {author} {\bibfnamefont {S.}~\bibnamefont
  {Succi}}\ and\ \bibinfo {author} {\bibfnamefont {R.}~\bibnamefont {Benzi}},\
  }\bibfield  {title} {\bibinfo {title} {Lattice boltzmann equation for quantum
  mechanics},\ }\href
  {https://doi.org/https://doi.org/10.1016/0167-2789(93)90096-J} {\bibfield
  {journal} {\bibinfo  {journal} {Physica D: Nonlinear Phenomena}\ }\textbf
  {\bibinfo {volume} {69}},\ \bibinfo {pages} {327} (\bibinfo {year}
  {1993})}\BibitemShut {NoStop}%
\bibitem [{\citenamefont {Fillion-Gourdeau}\ \emph {et~al.}(2012)\citenamefont
  {Fillion-Gourdeau}, \citenamefont {Lorin},\ and\ \citenamefont
  {Bandrauk}}]{FILLIONGOURDEAU20121403}%
  \BibitemOpen
  \bibfield  {author} {\bibinfo {author} {\bibfnamefont {F.}~\bibnamefont
  {Fillion-Gourdeau}}, \bibinfo {author} {\bibfnamefont {E.}~\bibnamefont
  {Lorin}},\ and\ \bibinfo {author} {\bibfnamefont {A.~D.}\ \bibnamefont
  {Bandrauk}},\ }\bibfield  {title} {\bibinfo {title} {Numerical solution of
  the time-dependent dirac equation in coordinate space without
  fermion-doubling},\ }\href
  {https://doi.org/https://doi.org/10.1016/j.cpc.2012.02.012} {\bibfield
  {journal} {\bibinfo  {journal} {Computer Physics Communications}\ }\textbf
  {\bibinfo {volume} {183}},\ \bibinfo {pages} {1403} (\bibinfo {year}
  {2012})}\BibitemShut {NoStop}%
\bibitem [{\citenamefont {Bao}\ and\ \citenamefont {Li}(2004)}]{BAO2004663}%
  \BibitemOpen
  \bibfield  {author} {\bibinfo {author} {\bibfnamefont {W.}~\bibnamefont
  {Bao}}\ and\ \bibinfo {author} {\bibfnamefont {X.-G.}\ \bibnamefont {Li}},\
  }\bibfield  {title} {\bibinfo {title} {An efficient and stable numerical
  method for the maxwell–dirac system},\ }\href
  {https://doi.org/https://doi.org/10.1016/j.jcp.2004.03.003} {\bibfield
  {journal} {\bibinfo  {journal} {Journal of Computational Physics}\ }\textbf
  {\bibinfo {volume} {199}},\ \bibinfo {pages} {663} (\bibinfo {year}
  {2004})}\BibitemShut {NoStop}%
\bibitem [{\citenamefont {Hammer}\ \emph {et~al.}(2014)\citenamefont {Hammer},
  \citenamefont {Pötz},\ and\ \citenamefont {Arnold}}]{HAMMER2014728}%
  \BibitemOpen
  \bibfield  {author} {\bibinfo {author} {\bibfnamefont {R.}~\bibnamefont
  {Hammer}}, \bibinfo {author} {\bibfnamefont {W.}~\bibnamefont {Pötz}},\ and\
  \bibinfo {author} {\bibfnamefont {A.}~\bibnamefont {Arnold}},\ }\bibfield
  {title} {\bibinfo {title} {A dispersion and norm preserving finite difference
  scheme with transparent boundary conditions for the dirac equation in
  (1+1)d},\ }\href {https://doi.org/https://doi.org/10.1016/j.jcp.2013.09.022}
  {\bibfield  {journal} {\bibinfo  {journal} {Journal of Computational
  Physics}\ }\textbf {\bibinfo {volume} {256}},\ \bibinfo {pages} {728}
  (\bibinfo {year} {2014})}\BibitemShut {NoStop}%
\bibitem [{\citenamefont {Hammer}\ and\ \citenamefont
  {Pötz}(2014)}]{HAMMER201440}%
  \BibitemOpen
  \bibfield  {author} {\bibinfo {author} {\bibfnamefont {R.}~\bibnamefont
  {Hammer}}\ and\ \bibinfo {author} {\bibfnamefont {W.}~\bibnamefont {Pötz}},\
  }\bibfield  {title} {\bibinfo {title} {Staggered grid leap-frog scheme for
  the (2+1)d dirac equation},\ }\href
  {https://doi.org/https://doi.org/10.1016/j.cpc.2013.08.013} {\bibfield
  {journal} {\bibinfo  {journal} {Computer Physics Communications}\ }\textbf
  {\bibinfo {volume} {185}},\ \bibinfo {pages} {40} (\bibinfo {year}
  {2014})}\BibitemShut {NoStop}%
\bibitem [{\citenamefont {P\"otz}(2017)}]{PhysRevE.96.053312}%
  \BibitemOpen
  \bibfield  {author} {\bibinfo {author} {\bibfnamefont {W.}~\bibnamefont
  {P\"otz}},\ }\bibfield  {title} {\bibinfo {title} {Single-cone
  finite-difference schemes for the (2+1)-dimensional dirac equation in general
  electromagnetic textures},\ }\href
  {https://doi.org/10.1103/PhysRevE.96.053312} {\bibfield  {journal} {\bibinfo
  {journal} {Phys. Rev. E}\ }\textbf {\bibinfo {volume} {96}},\ \bibinfo
  {pages} {053312} (\bibinfo {year} {2017})}\BibitemShut {NoStop}%
\bibitem [{\citenamefont {Antoine}\ \emph {et~al.}(2020)\citenamefont
  {Antoine}, \citenamefont {Fillion-Gourdeau}, \citenamefont {Lorin},\ and\
  \citenamefont {MacLean}}]{ANTOINE2020109412}%
  \BibitemOpen
  \bibfield  {author} {\bibinfo {author} {\bibfnamefont {X.}~\bibnamefont
  {Antoine}}, \bibinfo {author} {\bibfnamefont {F.}~\bibnamefont
  {Fillion-Gourdeau}}, \bibinfo {author} {\bibfnamefont {E.}~\bibnamefont
  {Lorin}},\ and\ \bibinfo {author} {\bibfnamefont {S.}~\bibnamefont
  {MacLean}},\ }\bibfield  {title} {\bibinfo {title} {Pseudospectral
  computational methods for the time-dependent dirac equation in static curved
  spaces},\ }\href {https://doi.org/https://doi.org/10.1016/j.jcp.2020.109412}
  {\bibfield  {journal} {\bibinfo  {journal} {Journal of Computational
  Physics}\ }\textbf {\bibinfo {volume} {411}},\ \bibinfo {pages} {109412}
  (\bibinfo {year} {2020})}\BibitemShut {NoStop}%
\bibitem [{\citenamefont {Fillion-Gourdeau}\ \emph {et~al.}(2021)\citenamefont
  {Fillion-Gourdeau}, \citenamefont {Lorin},\ and\ \citenamefont
  {MacLean}}]{PhysRevE.103.013312}%
  \BibitemOpen
  \bibfield  {author} {\bibinfo {author} {\bibfnamefont {F.}~\bibnamefont
  {Fillion-Gourdeau}}, \bibinfo {author} {\bibfnamefont {E.}~\bibnamefont
  {Lorin}},\ and\ \bibinfo {author} {\bibfnamefont {S.}~\bibnamefont
  {MacLean}},\ }\bibfield  {title} {\bibinfo {title} {Numerical quasiconformal
  transformations for electron dynamics on strained graphene surfaces},\ }\href
  {https://doi.org/10.1103/PhysRevE.103.013312} {\bibfield  {journal} {\bibinfo
   {journal} {Phys. Rev. E}\ }\textbf {\bibinfo {volume} {103}},\ \bibinfo
  {pages} {013312} (\bibinfo {year} {2021})}\BibitemShut {NoStop}%
\bibitem [{\citenamefont {Flouris}\ \emph {et~al.}(2018)\citenamefont
  {Flouris}, \citenamefont {Mendoza~Jimenez}, \citenamefont {Debus},\ and\
  \citenamefont {Herrmann}}]{PhysRevB.98.155419}%
  \BibitemOpen
  \bibfield  {author} {\bibinfo {author} {\bibfnamefont {K.}~\bibnamefont
  {Flouris}}, \bibinfo {author} {\bibfnamefont {M.}~\bibnamefont
  {Mendoza~Jimenez}}, \bibinfo {author} {\bibfnamefont {J.-D.}\ \bibnamefont
  {Debus}},\ and\ \bibinfo {author} {\bibfnamefont {H.~J.}\ \bibnamefont
  {Herrmann}},\ }\bibfield  {title} {\bibinfo {title} {Confining massless dirac
  particles in two-dimensional curved space},\ }\href
  {https://doi.org/10.1103/PhysRevB.98.155419} {\bibfield  {journal} {\bibinfo
  {journal} {Phys. Rev. B}\ }\textbf {\bibinfo {volume} {98}},\ \bibinfo
  {pages} {155419} (\bibinfo {year} {2018})}\BibitemShut {NoStop}%
\bibitem [{\citenamefont {Maslov}\ and\ \citenamefont
  {Fedoriuk}(1981)}]{maslov2001semi}%
  \BibitemOpen
  \bibfield  {author} {\bibinfo {author} {\bibfnamefont {V.~P.}\ \bibnamefont
  {Maslov}}\ and\ \bibinfo {author} {\bibfnamefont {M.~V.}\ \bibnamefont
  {Fedoriuk}},\ }\href@noop {} {\emph {\bibinfo {title} {Semi-classical
  approximation in quantum mechanics}}},\ Vol.~\bibinfo {volume} {7}\ (\bibinfo
   {publisher} {D. Reidel Publishing Company},\ \bibinfo {address} {Dordrecht,
  Holland},\ \bibinfo {year} {1981})\BibitemShut {NoStop}%
\bibitem [{\citenamefont {Bolte}\ and\ \citenamefont
  {Keppeler}(1999)}]{BOLTE1999125}%
  \BibitemOpen
  \bibfield  {author} {\bibinfo {author} {\bibfnamefont {J.}~\bibnamefont
  {Bolte}}\ and\ \bibinfo {author} {\bibfnamefont {S.}~\bibnamefont
  {Keppeler}},\ }\bibfield  {title} {\bibinfo {title} {A semiclassical approach
  to the dirac equation},\ }\href
  {https://doi.org/https://doi.org/10.1006/aphy.1999.5912} {\bibfield
  {journal} {\bibinfo  {journal} {Annals of Physics}\ }\textbf {\bibinfo
  {volume} {274}},\ \bibinfo {pages} {125} (\bibinfo {year}
  {1999})}\BibitemShut {NoStop}%
\bibitem [{\citenamefont {Spohn}(2000)}]{SPOHN2000420}%
  \BibitemOpen
  \bibfield  {author} {\bibinfo {author} {\bibfnamefont {H.}~\bibnamefont
  {Spohn}},\ }\bibfield  {title} {\bibinfo {title} {Semiclassical limit of the
  dirac equation and spin precession},\ }\href
  {https://doi.org/https://doi.org/10.1006/aphy.2000.6039} {\bibfield
  {journal} {\bibinfo  {journal} {Annals of Physics}\ }\textbf {\bibinfo
  {volume} {282}},\ \bibinfo {pages} {420} (\bibinfo {year}
  {2000})}\BibitemShut {NoStop}%
\bibitem [{\citenamefont {Sparber}\ and\ \citenamefont
  {Markowich}(2003)}]{doi:10.1063/1.1604455}%
  \BibitemOpen
  \bibfield  {author} {\bibinfo {author} {\bibfnamefont {C.}~\bibnamefont
  {Sparber}}\ and\ \bibinfo {author} {\bibfnamefont {P.}~\bibnamefont
  {Markowich}},\ }\bibfield  {title} {\bibinfo {title} {Semiclassical
  asymptotics for the maxwell-dirac system},\ }\href
  {https://doi.org/10.1063/1.1604455} {\bibfield  {journal} {\bibinfo
  {journal} {Journal of Mathematical Physics}\ }\textbf {\bibinfo {volume}
  {44}},\ \bibinfo {pages} {4555} (\bibinfo {year} {2003})}\BibitemShut
  {NoStop}%
\bibitem [{\citenamefont {Reijnders}\ \emph {et~al.}(2013)\citenamefont
  {Reijnders}, \citenamefont {Tudorovskiy},\ and\ \citenamefont
  {Katsnelson}}]{REIJNDERS2013155}%
  \BibitemOpen
  \bibfield  {author} {\bibinfo {author} {\bibfnamefont {K.}~\bibnamefont
  {Reijnders}}, \bibinfo {author} {\bibfnamefont {T.}~\bibnamefont
  {Tudorovskiy}},\ and\ \bibinfo {author} {\bibfnamefont {M.}~\bibnamefont
  {Katsnelson}},\ }\bibfield  {title} {\bibinfo {title} {Semiclassical theory
  of potential scattering for massless dirac fermions},\ }\href
  {https://doi.org/https://doi.org/10.1016/j.aop.2013.03.001} {\bibfield
  {journal} {\bibinfo  {journal} {Annals of Physics}\ }\textbf {\bibinfo
  {volume} {333}},\ \bibinfo {pages} {155} (\bibinfo {year}
  {2013})}\BibitemShut {NoStop}%
\bibitem [{\citenamefont {Bolte}\ and\ \citenamefont
  {Keppeler}(1998)}]{PhysRevLett.81.1987}%
  \BibitemOpen
  \bibfield  {author} {\bibinfo {author} {\bibfnamefont {J.}~\bibnamefont
  {Bolte}}\ and\ \bibinfo {author} {\bibfnamefont {S.}~\bibnamefont
  {Keppeler}},\ }\bibfield  {title} {\bibinfo {title} {Semiclassical time
  evolution and trace formula for relativistic spin-1/2 particles},\ }\href
  {https://doi.org/10.1103/PhysRevLett.81.1987} {\bibfield  {journal} {\bibinfo
   {journal} {Phys. Rev. Lett.}\ }\textbf {\bibinfo {volume} {81}},\ \bibinfo
  {pages} {1987} (\bibinfo {year} {1998})}\BibitemShut {NoStop}%
\bibitem [{\citenamefont {B{\o}ggild}\ \emph {et~al.}(2017)\citenamefont
  {B{\o}ggild}, \citenamefont {Caridad}, \citenamefont {Stampfer},
  \citenamefont {Calogero}, \citenamefont {Papior},\ and\ \citenamefont
  {Brandbyge}}]{boggild2017two}%
  \BibitemOpen
  \bibfield  {author} {\bibinfo {author} {\bibfnamefont {P.}~\bibnamefont
  {B{\o}ggild}}, \bibinfo {author} {\bibfnamefont {J.~M.}\ \bibnamefont
  {Caridad}}, \bibinfo {author} {\bibfnamefont {C.}~\bibnamefont {Stampfer}},
  \bibinfo {author} {\bibfnamefont {G.}~\bibnamefont {Calogero}}, \bibinfo
  {author} {\bibfnamefont {N.~R.}\ \bibnamefont {Papior}},\ and\ \bibinfo
  {author} {\bibfnamefont {M.}~\bibnamefont {Brandbyge}},\ }\bibfield  {title}
  {\bibinfo {title} {A two-dimensional dirac fermion microscope},\ }\href
  {https://doi.org/https://doi.org/10.1038/ncomms15783} {\bibfield  {journal}
  {\bibinfo  {journal} {Nature communications}\ }\textbf {\bibinfo {volume}
  {8}},\ \bibinfo {pages} {1} (\bibinfo {year} {2017})}\BibitemShut {NoStop}%
\bibitem [{\citenamefont {Carmier}\ and\ \citenamefont
  {Ullmo}(2008)}]{PhysRevB.77.245413}%
  \BibitemOpen
  \bibfield  {author} {\bibinfo {author} {\bibfnamefont {P.}~\bibnamefont
  {Carmier}}\ and\ \bibinfo {author} {\bibfnamefont {D.}~\bibnamefont
  {Ullmo}},\ }\bibfield  {title} {\bibinfo {title} {Berry phase in graphene:
  Semiclassical perspective},\ }\href
  {https://doi.org/10.1103/PhysRevB.77.245413} {\bibfield  {journal} {\bibinfo
  {journal} {Phys. Rev. B}\ }\textbf {\bibinfo {volume} {77}},\ \bibinfo
  {pages} {245413} (\bibinfo {year} {2008})}\BibitemShut {NoStop}%
\bibitem [{\citenamefont {Reijnders}\ \emph {et~al.}(2018)\citenamefont
  {Reijnders}, \citenamefont {Minenkov}, \citenamefont {Katsnelson},\ and\
  \citenamefont {Dobrokhotov}}]{REIJNDERS201865}%
  \BibitemOpen
  \bibfield  {author} {\bibinfo {author} {\bibfnamefont {K.}~\bibnamefont
  {Reijnders}}, \bibinfo {author} {\bibfnamefont {D.}~\bibnamefont {Minenkov}},
  \bibinfo {author} {\bibfnamefont {M.}~\bibnamefont {Katsnelson}},\ and\
  \bibinfo {author} {\bibfnamefont {S.}~\bibnamefont {Dobrokhotov}},\
  }\bibfield  {title} {\bibinfo {title} {Electronic optics in graphene in the
  semiclassical approximation},\ }\href
  {https://doi.org/https://doi.org/10.1016/j.aop.2018.08.004} {\bibfield
  {journal} {\bibinfo  {journal} {Annals of Physics}\ }\textbf {\bibinfo
  {volume} {397}},\ \bibinfo {pages} {65} (\bibinfo {year} {2018})}\BibitemShut
  {NoStop}%
\bibitem [{\citenamefont {Paredes-Rocha}\ \emph {et~al.}(2021)\citenamefont
  {Paredes-Rocha}, \citenamefont {Betancur-Ocampo}, \citenamefont {Szpak},\
  and\ \citenamefont {Stegmann}}]{PhysRevB.103.045404}%
  \BibitemOpen
  \bibfield  {author} {\bibinfo {author} {\bibfnamefont {E.}~\bibnamefont
  {Paredes-Rocha}}, \bibinfo {author} {\bibfnamefont {Y.}~\bibnamefont
  {Betancur-Ocampo}}, \bibinfo {author} {\bibfnamefont {N.}~\bibnamefont
  {Szpak}},\ and\ \bibinfo {author} {\bibfnamefont {T.}~\bibnamefont
  {Stegmann}},\ }\bibfield  {title} {\bibinfo {title} {Gradient-index electron
  optics in graphene $p\ensuremath{-}n$ junctions},\ }\href
  {https://doi.org/10.1103/PhysRevB.103.045404} {\bibfield  {journal} {\bibinfo
   {journal} {Phys. Rev. B}\ }\textbf {\bibinfo {volume} {103}},\ \bibinfo
  {pages} {045404} (\bibinfo {year} {2021})}\BibitemShut {NoStop}%
\bibitem [{\citenamefont {Reijnders}\ and\ \citenamefont
  {Katsnelson}(2017)}]{PhysRevB.96.045305}%
  \BibitemOpen
  \bibfield  {author} {\bibinfo {author} {\bibfnamefont {K.~J.~A.}\
  \bibnamefont {Reijnders}}\ and\ \bibinfo {author} {\bibfnamefont {M.~I.}\
  \bibnamefont {Katsnelson}},\ }\bibfield  {title} {\bibinfo {title}
  {Diffraction catastrophes and semiclassical quantum mechanics for veselago
  lensing in graphene},\ }\href {https://doi.org/10.1103/PhysRevB.96.045305}
  {\bibfield  {journal} {\bibinfo  {journal} {Phys. Rev. B}\ }\textbf {\bibinfo
  {volume} {96}},\ \bibinfo {pages} {045305} (\bibinfo {year}
  {2017})}\BibitemShut {NoStop}%
\bibitem [{\citenamefont {Silenko}\ and\ \citenamefont
  {Teryaev}(2005)}]{PhysRevD.71.064016}%
  \BibitemOpen
  \bibfield  {author} {\bibinfo {author} {\bibfnamefont {A.~J.}\ \bibnamefont
  {Silenko}}\ and\ \bibinfo {author} {\bibfnamefont {O.~V.}\ \bibnamefont
  {Teryaev}},\ }\bibfield  {title} {\bibinfo {title} {Semiclassical limit for
  dirac particles interacting with a gravitational field},\ }\href
  {https://doi.org/10.1103/PhysRevD.71.064016} {\bibfield  {journal} {\bibinfo
  {journal} {Phys. Rev. D}\ }\textbf {\bibinfo {volume} {71}},\ \bibinfo
  {pages} {064016} (\bibinfo {year} {2005})}\BibitemShut {NoStop}%
\bibitem [{\citenamefont {Gosselin}\ \emph {et~al.}(2007)\citenamefont
  {Gosselin}, \citenamefont {Bérard},\ and\ \citenamefont
  {Mohrbach}}]{GOSSELIN2007356}%
  \BibitemOpen
  \bibfield  {author} {\bibinfo {author} {\bibfnamefont {P.}~\bibnamefont
  {Gosselin}}, \bibinfo {author} {\bibfnamefont {A.}~\bibnamefont {Bérard}},\
  and\ \bibinfo {author} {\bibfnamefont {H.}~\bibnamefont {Mohrbach}},\
  }\bibfield  {title} {\bibinfo {title} {Semiclassical dynamics of dirac
  particles interacting with a static gravitational field},\ }\href
  {https://doi.org/https://doi.org/10.1016/j.physleta.2007.04.022} {\bibfield
  {journal} {\bibinfo  {journal} {Physics Letters A}\ }\textbf {\bibinfo
  {volume} {368}},\ \bibinfo {pages} {356} (\bibinfo {year}
  {2007})}\BibitemShut {NoStop}%
\bibitem [{\citenamefont {Cianfrani}\ and\ \citenamefont
  {Montani}(2008{\natexlab{a}})}]{doi:10.1142/S0217751X08040214}%
  \BibitemOpen
  \bibfield  {author} {\bibinfo {author} {\bibfnamefont {F.}~\bibnamefont
  {Cianfrani}}\ and\ \bibinfo {author} {\bibfnamefont {G.}~\bibnamefont
  {Montani}},\ }\bibfield  {title} {\bibinfo {title} {Curvature-spin coupling
  from the semi-classical limit of the dirac equation},\ }\href
  {https://doi.org/10.1142/S0217751X08040214} {\bibfield  {journal} {\bibinfo
  {journal} {International Journal of Modern Physics A}\ }\textbf {\bibinfo
  {volume} {23}},\ \bibinfo {pages} {1274} (\bibinfo {year}
  {2008}{\natexlab{a}})},\ \Eprint
  {https://arxiv.org/abs/https://doi.org/10.1142/S0217751X08040214}
  {https://doi.org/10.1142/S0217751X08040214} \BibitemShut {NoStop}%
\bibitem [{\citenamefont {Cianfrani}\ and\ \citenamefont
  {Montani}(2008{\natexlab{b}})}]{Montani2008}%
  \BibitemOpen
  \bibfield  {author} {\bibinfo {author} {\bibfnamefont {F.}~\bibnamefont
  {Cianfrani}}\ and\ \bibinfo {author} {\bibfnamefont {G.}~\bibnamefont
  {Montani}},\ }\bibfield  {title} {\bibinfo {title} {Dirac equations in curved
  space-time vs. papapetrou spinning particles},\ }\href
  {https://doi.org/10.1209/0295-5075/84/30008} {\ \textbf {\bibinfo {volume}
  {84}},\ \bibinfo {pages} {30008} (\bibinfo {year}
  {2008}{\natexlab{b}})}\BibitemShut {NoStop}%
\bibitem [{\citenamefont {de~Juan}\ \emph {et~al.}(2013)\citenamefont
  {de~Juan}, \citenamefont {Ma\~nes},\ and\ \citenamefont
  {Vozmediano}}]{PhysRevB.87.165131}%
  \BibitemOpen
  \bibfield  {author} {\bibinfo {author} {\bibfnamefont {F.}~\bibnamefont
  {de~Juan}}, \bibinfo {author} {\bibfnamefont {J.~L.}\ \bibnamefont
  {Ma\~nes}},\ and\ \bibinfo {author} {\bibfnamefont {M.~A.~H.}\ \bibnamefont
  {Vozmediano}},\ }\bibfield  {title} {\bibinfo {title} {Gauge fields from
  strain in graphene},\ }\href {https://doi.org/10.1103/PhysRevB.87.165131}
  {\bibfield  {journal} {\bibinfo  {journal} {Phys. Rev. B}\ }\textbf {\bibinfo
  {volume} {87}},\ \bibinfo {pages} {165131} (\bibinfo {year}
  {2013})}\BibitemShut {NoStop}%
\bibitem [{\citenamefont {Vozmediano}\ \emph {et~al.}(2010)\citenamefont
  {Vozmediano}, \citenamefont {Katsnelson},\ and\ \citenamefont
  {Guinea}}]{VOZMEDIANO2010109}%
  \BibitemOpen
  \bibfield  {author} {\bibinfo {author} {\bibfnamefont {M.}~\bibnamefont
  {Vozmediano}}, \bibinfo {author} {\bibfnamefont {M.}~\bibnamefont
  {Katsnelson}},\ and\ \bibinfo {author} {\bibfnamefont {F.}~\bibnamefont
  {Guinea}},\ }\bibfield  {title} {\bibinfo {title} {Gauge fields in
  graphene},\ }\href
  {https://doi.org/https://doi.org/10.1016/j.physrep.2010.07.003} {\bibfield
  {journal} {\bibinfo  {journal} {Physics Reports}\ }\textbf {\bibinfo {volume}
  {496}},\ \bibinfo {pages} {109 } (\bibinfo {year} {2010})}\BibitemShut
  {NoStop}%
\bibitem [{\citenamefont {de~Juan}\ \emph {et~al.}(2012)\citenamefont
  {de~Juan}, \citenamefont {Sturla},\ and\ \citenamefont
  {Vozmediano}}]{PhysRevLett.108.227205}%
  \BibitemOpen
  \bibfield  {author} {\bibinfo {author} {\bibfnamefont {F.}~\bibnamefont
  {de~Juan}}, \bibinfo {author} {\bibfnamefont {M.}~\bibnamefont {Sturla}},\
  and\ \bibinfo {author} {\bibfnamefont {M.~A.~H.}\ \bibnamefont
  {Vozmediano}},\ }\bibfield  {title} {\bibinfo {title} {Space dependent fermi
  velocity in strained graphene},\ }\href
  {https://doi.org/10.1103/PhysRevLett.108.227205} {\bibfield  {journal}
  {\bibinfo  {journal} {Phys. Rev. Lett.}\ }\textbf {\bibinfo {volume} {108}},\
  \bibinfo {pages} {227205} (\bibinfo {year} {2012})}\BibitemShut {NoStop}%
\bibitem [{\citenamefont {Volovik}\ and\ \citenamefont
  {Zubkov}(2014)}]{VOLOVIK2014352}%
  \BibitemOpen
  \bibfield  {author} {\bibinfo {author} {\bibfnamefont {G.}~\bibnamefont
  {Volovik}}\ and\ \bibinfo {author} {\bibfnamefont {M.}~\bibnamefont
  {Zubkov}},\ }\bibfield  {title} {\bibinfo {title} {Emergent horava gravity in
  graphene},\ }\href
  {https://doi.org/https://doi.org/10.1016/j.aop.2013.11.003} {\bibfield
  {journal} {\bibinfo  {journal} {Annals of Physics}\ }\textbf {\bibinfo
  {volume} {340}},\ \bibinfo {pages} {352 } (\bibinfo {year}
  {2014})}\BibitemShut {NoStop}%
\bibitem [{\citenamefont {Amorim}\ \emph {et~al.}(2016)\citenamefont {Amorim},
  \citenamefont {Cortijo}, \citenamefont {[de Juan]}, \citenamefont {Grushin},
  \citenamefont {Guinea}, \citenamefont {Gutiérrez-Rubio}, \citenamefont
  {Ochoa}, \citenamefont {Parente}, \citenamefont {Roldán}, \citenamefont
  {San-Jose}, \citenamefont {Schiefele}, \citenamefont {Sturla},\ and\
  \citenamefont {Vozmediano}}]{AMORIM20161}%
  \BibitemOpen
  \bibfield  {author} {\bibinfo {author} {\bibfnamefont {B.}~\bibnamefont
  {Amorim}}, \bibinfo {author} {\bibfnamefont {A.}~\bibnamefont {Cortijo}},
  \bibinfo {author} {\bibfnamefont {F.}~\bibnamefont {[de Juan]}}, \bibinfo
  {author} {\bibfnamefont {A.}~\bibnamefont {Grushin}}, \bibinfo {author}
  {\bibfnamefont {F.}~\bibnamefont {Guinea}}, \bibinfo {author} {\bibfnamefont
  {A.}~\bibnamefont {Gutiérrez-Rubio}}, \bibinfo {author} {\bibfnamefont
  {H.}~\bibnamefont {Ochoa}}, \bibinfo {author} {\bibfnamefont
  {V.}~\bibnamefont {Parente}}, \bibinfo {author} {\bibfnamefont
  {R.}~\bibnamefont {Roldán}}, \bibinfo {author} {\bibfnamefont
  {P.}~\bibnamefont {San-Jose}}, \bibinfo {author} {\bibfnamefont
  {J.}~\bibnamefont {Schiefele}}, \bibinfo {author} {\bibfnamefont
  {M.}~\bibnamefont {Sturla}},\ and\ \bibinfo {author} {\bibfnamefont
  {M.}~\bibnamefont {Vozmediano}},\ }\bibfield  {title} {\bibinfo {title}
  {Novel effects of strains in graphene and other two dimensional materials},\
  }\href {https://doi.org/https://doi.org/10.1016/j.physrep.2015.12.006}
  {\bibfield  {journal} {\bibinfo  {journal} {Physics Reports}\ }\textbf
  {\bibinfo {volume} {617}},\ \bibinfo {pages} {1 } (\bibinfo {year}
  {2016})}\BibitemShut {NoStop}%
\bibitem [{\citenamefont {Oliva-Leyva}\ and\ \citenamefont
  {Naumis}(2015)}]{OLIVALEYVA20152645}%
  \BibitemOpen
  \bibfield  {author} {\bibinfo {author} {\bibfnamefont {M.}~\bibnamefont
  {Oliva-Leyva}}\ and\ \bibinfo {author} {\bibfnamefont {G.~G.}\ \bibnamefont
  {Naumis}},\ }\bibfield  {title} {\bibinfo {title} {Generalizing the fermi
  velocity of strained graphene from uniform to nonuniform strain},\ }\href
  {https://doi.org/https://doi.org/10.1016/j.physleta.2015.05.039} {\bibfield
  {journal} {\bibinfo  {journal} {Physics Letters A}\ }\textbf {\bibinfo
  {volume} {379}},\ \bibinfo {pages} {2645} (\bibinfo {year}
  {2015})}\BibitemShut {NoStop}%
\bibitem [{\citenamefont {Naumis}\ \emph {et~al.}(2017)\citenamefont {Naumis},
  \citenamefont {Barraza-Lopez}, \citenamefont {Oliva-Leyva},\ and\
  \citenamefont {Terrones}}]{Naumis_2017}%
  \BibitemOpen
  \bibfield  {author} {\bibinfo {author} {\bibfnamefont {G.~G.}\ \bibnamefont
  {Naumis}}, \bibinfo {author} {\bibfnamefont {S.}~\bibnamefont
  {Barraza-Lopez}}, \bibinfo {author} {\bibfnamefont {M.}~\bibnamefont
  {Oliva-Leyva}},\ and\ \bibinfo {author} {\bibfnamefont {H.}~\bibnamefont
  {Terrones}},\ }\bibfield  {title} {\bibinfo {title} {Electronic and optical
  properties of strained graphene and other strained 2d materials: a review},\
  }\href {https://doi.org/10.1088/1361-6633/aa74ef} {\bibfield  {journal}
  {\bibinfo  {journal} {Reports on Progress in Physics}\ }\textbf {\bibinfo
  {volume} {80}},\ \bibinfo {pages} {096501} (\bibinfo {year}
  {2017})}\BibitemShut {NoStop}%
\bibitem [{\citenamefont {Stegmann}\ and\ \citenamefont
  {Szpak}(2016)}]{Stegmann_2016}%
  \BibitemOpen
  \bibfield  {author} {\bibinfo {author} {\bibfnamefont {T.}~\bibnamefont
  {Stegmann}}\ and\ \bibinfo {author} {\bibfnamefont {N.}~\bibnamefont
  {Szpak}},\ }\bibfield  {title} {\bibinfo {title} {Current flow paths in
  deformed graphene: from quantum transport to classical trajectories in curved
  space},\ }\href {https://doi.org/10.1088/1367-2630/18/5/053016} {\bibfield
  {journal} {\bibinfo  {journal} {New Journal of Physics}\ }\textbf {\bibinfo
  {volume} {18}},\ \bibinfo {pages} {053016} (\bibinfo {year}
  {2016})}\BibitemShut {NoStop}%
\bibitem [{\citenamefont {Pollock}(2010)}]{pollock2010dirac}%
  \BibitemOpen
  \bibfield  {author} {\bibinfo {author} {\bibfnamefont {M.}~\bibnamefont
  {Pollock}},\ }\bibfield  {title} {\bibinfo {title} {On the dirac equation in
  curved space-time.},\ }\href@noop {} {\bibfield  {journal} {\bibinfo
  {journal} {Acta Physica Polonica B}\ }\textbf {\bibinfo {volume} {41}}
  (\bibinfo {year} {2010})}\BibitemShut {NoStop}%
\bibitem [{\citenamefont {Arias}\ \emph {et~al.}(2015)\citenamefont {Arias},
  \citenamefont {Hern\'andez},\ and\ \citenamefont
  {Lewenkopf}}]{PhysRevB.92.245110}%
  \BibitemOpen
  \bibfield  {author} {\bibinfo {author} {\bibfnamefont {E.}~\bibnamefont
  {Arias}}, \bibinfo {author} {\bibfnamefont {A.~R.}\ \bibnamefont
  {Hern\'andez}},\ and\ \bibinfo {author} {\bibfnamefont {C.}~\bibnamefont
  {Lewenkopf}},\ }\bibfield  {title} {\bibinfo {title} {Gauge fields in
  graphene with nonuniform elastic deformations: A quantum field theory
  approach},\ }\href {https://doi.org/10.1103/PhysRevB.92.245110} {\bibfield
  {journal} {\bibinfo  {journal} {Phys. Rev. B}\ }\textbf {\bibinfo {volume}
  {92}},\ \bibinfo {pages} {245110} (\bibinfo {year} {2015})}\BibitemShut
  {NoStop}%
\bibitem [{\citenamefont {Iorio}\ and\ \citenamefont
  {Pais}(2015)}]{PhysRevD.92.125005}%
  \BibitemOpen
  \bibfield  {author} {\bibinfo {author} {\bibfnamefont {A.}~\bibnamefont
  {Iorio}}\ and\ \bibinfo {author} {\bibfnamefont {P.}~\bibnamefont {Pais}},\
  }\bibfield  {title} {\bibinfo {title} {Revisiting the gauge fields of
  strained graphene},\ }\href {https://doi.org/10.1103/PhysRevD.92.125005}
  {\bibfield  {journal} {\bibinfo  {journal} {Phys. Rev. D}\ }\textbf {\bibinfo
  {volume} {92}},\ \bibinfo {pages} {125005} (\bibinfo {year}
  {2015})}\BibitemShut {NoStop}%
\bibitem [{\citenamefont {Ahlfors}(2006)}]{ahlfors2006lectures}%
  \BibitemOpen
  \bibfield  {author} {\bibinfo {author} {\bibfnamefont {L.~V.}\ \bibnamefont
  {Ahlfors}},\ }\href@noop {} {\emph {\bibinfo {title} {Lectures on
  quasiconformal mappings}}},\ Vol.~\bibinfo {volume} {38}\ (\bibinfo
  {publisher} {American Mathematical Soc.},\ \bibinfo {year}
  {2006})\BibitemShut {NoStop}%
\bibitem [{\citenamefont {Qiu}\ \emph {et~al.}(2019)\citenamefont {Qiu},
  \citenamefont {Lam},\ and\ \citenamefont {Lui}}]{origami}%
  \BibitemOpen
  \bibfield  {author} {\bibinfo {author} {\bibfnamefont {D.}~\bibnamefont
  {Qiu}}, \bibinfo {author} {\bibfnamefont {K.-C.}\ \bibnamefont {Lam}},\ and\
  \bibinfo {author} {\bibfnamefont {L.-M.}\ \bibnamefont {Lui}},\ }\bibfield
  {title} {\bibinfo {title} {Computing quasi-conformal folds},\ }\href@noop {}
  {\bibfield  {journal} {\bibinfo  {journal} {SIAM Journal on Imaging
  Sciences}\ }\textbf {\bibinfo {volume} {12}},\ \bibinfo {pages} {1392}
  (\bibinfo {year} {2019})}\BibitemShut {NoStop}%
\bibitem [{\citenamefont {Houchmandzadeh}(2020)}]{doi:10.1119/10.0000781}%
  \BibitemOpen
  \bibfield  {author} {\bibinfo {author} {\bibfnamefont {B.}~\bibnamefont
  {Houchmandzadeh}},\ }\bibfield  {title} {\bibinfo {title} {The
  hamilton–jacobi equation: An alternative approach},\ }\href
  {https://doi.org/10.1119/10.0000781} {\bibfield  {journal} {\bibinfo
  {journal} {American Journal of Physics}\ }\textbf {\bibinfo {volume} {88}},\
  \bibinfo {pages} {353} (\bibinfo {year} {2020})}\BibitemShut {NoStop}%
\bibitem [{\citenamefont {Evans}\ and\ \citenamefont
  {Rosenquist}(1986)}]{doi:10.1119/1.14861}%
  \BibitemOpen
  \bibfield  {author} {\bibinfo {author} {\bibfnamefont {J.}~\bibnamefont
  {Evans}}\ and\ \bibinfo {author} {\bibfnamefont {M.}~\bibnamefont
  {Rosenquist}},\ }\bibfield  {title} {\bibinfo {title} {‘‘f=ma’’
  optics},\ }\href {https://doi.org/10.1119/1.14861} {\bibfield  {journal}
  {\bibinfo  {journal} {American Journal of Physics}\ }\textbf {\bibinfo
  {volume} {54}},\ \bibinfo {pages} {876} (\bibinfo {year} {1986})}\BibitemShut
  {NoStop}%
\bibitem [{\citenamefont {Lorin}\ \emph {et~al.}(2021)\citenamefont {Lorin},
  \citenamefont {Fillion-Gourdeau},\ and\ \citenamefont {MacLean}}]{lorin2021}%
  \BibitemOpen
  \bibfield  {author} {\bibinfo {author} {\bibfnamefont {E.}~\bibnamefont
  {Lorin}}, \bibinfo {author} {\bibfnamefont {F.}~\bibnamefont
  {Fillion-Gourdeau}},\ and\ \bibinfo {author} {\bibfnamefont {S.}~\bibnamefont
  {MacLean}},\ }\bibfield  {title} {\bibinfo {title} {Inverse design of
  strained graphene surfaces for electron control},\ }\href@noop {} {\
  (\bibinfo {year} {2021})},\ \bibinfo {note} {submitted to Computer Physics
  Communications}\BibitemShut {NoStop}%
\bibitem [{\citenamefont {Radhakrishnan}\ and\ \citenamefont
  {Hindmarsh}()}]{osti_15013302}%
  \BibitemOpen
  \bibfield  {author} {\bibinfo {author} {\bibfnamefont {K.}~\bibnamefont
  {Radhakrishnan}}\ and\ \bibinfo {author} {\bibfnamefont {A.~C.}\ \bibnamefont
  {Hindmarsh}},\ }\bibfield  {title} {\bibinfo {title} {Description and use of
  lsode, the livemore solver for ordinary differential equations}\ }\href
  {https://doi.org/10.2172/15013302} {10.2172/15013302}\BibitemShut {NoStop}%
\bibitem [{\citenamefont {Ye}\ and\ \citenamefont {Lin}(2008)}]{gravlens2008}%
  \BibitemOpen
  \bibfield  {author} {\bibinfo {author} {\bibfnamefont {X.-H.}\ \bibnamefont
  {Ye}}\ and\ \bibinfo {author} {\bibfnamefont {Q.}~\bibnamefont {Lin}},\
  }\bibfield  {title} {\bibinfo {title} {Gravitational lensing analysed by the
  graded refractive index of a vacuum},\ }\href
  {https://doi.org/10.1088/1464-4258/10/7/075001} {\ \textbf {\bibinfo {volume}
  {10}},\ \bibinfo {pages} {075001} (\bibinfo {year} {2008})}\BibitemShut
  {NoStop}%
\bibitem [{\citenamefont {Sakoda}(2004)}]{sakoda2004optical}%
  \BibitemOpen
  \bibfield  {author} {\bibinfo {author} {\bibfnamefont {K.}~\bibnamefont
  {Sakoda}},\ }\href@noop {} {\emph {\bibinfo {title} {Optical properties of
  photonic crystals}}},\ Vol.~\bibinfo {volume} {80}\ (\bibinfo  {publisher}
  {Springer Science \& Business Media},\ \bibinfo {year} {2004})\BibitemShut
  {NoStop}%
\bibitem [{\citenamefont {Huang}\ \emph {et~al.}(2012)\citenamefont {Huang},
  \citenamefont {Jin}, \citenamefont {Wu},\ and\ \citenamefont
  {Yin}}]{huang2012gaussian}%
  \BibitemOpen
  \bibfield  {author} {\bibinfo {author} {\bibfnamefont {Z.}~\bibnamefont
  {Huang}}, \bibinfo {author} {\bibfnamefont {S.}~\bibnamefont {Jin}}, \bibinfo
  {author} {\bibfnamefont {H.}~\bibnamefont {Wu}},\ and\ \bibinfo {author}
  {\bibfnamefont {D.}~\bibnamefont {Yin}},\ }\bibfield  {title} {\bibinfo
  {title} {Gaussian beam methods for the dirac equation in the semi-classical
  regime},\ }\href
  {https://doi.org/https://dx.doi.org/10.4310/CMS.2012.v10.n4.a14} {\bibfield
  {journal} {\bibinfo  {journal} {Communications in Mathematical Sciences}\
  }\textbf {\bibinfo {volume} {10}},\ \bibinfo {pages} {1301} (\bibinfo {year}
  {2012})}\BibitemShut {NoStop}%
\bibitem [{\citenamefont {Chai}\ \emph {et~al.}(2019)\citenamefont {Chai},
  \citenamefont {Lorin},\ and\ \citenamefont {Yang}}]{doi:10.1137/18M1222831}%
  \BibitemOpen
  \bibfield  {author} {\bibinfo {author} {\bibfnamefont {L.}~\bibnamefont
  {Chai}}, \bibinfo {author} {\bibfnamefont {E.}~\bibnamefont {Lorin}},\ and\
  \bibinfo {author} {\bibfnamefont {X.}~\bibnamefont {Yang}},\ }\bibfield
  {title} {\bibinfo {title} {Frozen gaussian approximation for the dirac
  equation in semiclassical regime},\ }\href
  {https://doi.org/10.1137/18M1222831} {\bibfield  {journal} {\bibinfo
  {journal} {SIAM Journal on Numerical Analysis}\ }\textbf {\bibinfo {volume}
  {57}},\ \bibinfo {pages} {2383} (\bibinfo {year} {2019})}\BibitemShut
  {NoStop}%
\bibitem [{\citenamefont {Parker}(1980)}]{PhysRevD.22.1922}%
  \BibitemOpen
  \bibfield  {author} {\bibinfo {author} {\bibfnamefont {L.}~\bibnamefont
  {Parker}},\ }\bibfield  {title} {\bibinfo {title} {One-electron atom as a
  probe of spacetime curvature},\ }\href
  {https://doi.org/10.1103/PhysRevD.22.1922} {\bibfield  {journal} {\bibinfo
  {journal} {Phys. Rev. D}\ }\textbf {\bibinfo {volume} {22}},\ \bibinfo
  {pages} {1922} (\bibinfo {year} {1980})}\BibitemShut {NoStop}%
\bibitem [{\citenamefont {Gorbatenko}\ and\ \citenamefont
  {Neznamov}(2011)}]{PhysRevD.83.105002}%
  \BibitemOpen
  \bibfield  {author} {\bibinfo {author} {\bibfnamefont {M.~V.}\ \bibnamefont
  {Gorbatenko}}\ and\ \bibinfo {author} {\bibfnamefont {V.~P.}\ \bibnamefont
  {Neznamov}},\ }\bibfield  {title} {\bibinfo {title} {Uniqueness and
  self-conjugacy of dirac hamiltonians in arbitrary gravitational fields},\
  }\href {https://doi.org/10.1103/PhysRevD.83.105002} {\bibfield  {journal}
  {\bibinfo  {journal} {Phys. Rev. D}\ }\textbf {\bibinfo {volume} {83}},\
  \bibinfo {pages} {105002} (\bibinfo {year} {2011})}\BibitemShut {NoStop}%
\end{thebibliography}%

\end{document}